\documentclass[aps,prb,superscriptaddress,reprint]{revtex4-2}
\usepackage{CJK}
\usepackage[pdftex]{graphicx,xcolor,hyperref}
\usepackage{amsmath,amssymb,bm}
\usepackage{array}

{\par}

\begin{document}
\title{Dynamic Magnetoelastic Boundary Conditions and the Pumping of Phonons}
\begin{CJK}{UTF8}{ipxm}
\author{Takuma \surname{Sato} (佐藤拓磨)}
\author{Weichao \surname{Yu} (\CJKfamily{gbsn}余伟超)}
\affiliation{Institute for Materials Research, Tohoku University, Sendai 980-8577, Japan}
\author{Simon \surname{Streib}}
\affiliation{Department of Physics and Astronomy, Uppsala University, Box 516, SE-75120 Uppsala, Sweden}
\author{Gerrit E. W. \surname{Bauer}}
\affiliation{Institute for Materials Research, Tohoku University, Sendai 980-8577, Japan}
\affiliation{WPI-AIMR, Tohoku University, Sendai 980-8577, Japan}
\affiliation{Zernike Institute for Advanced Materials, Groningen University, Groningen, The Netherlands}
\date{\today}
\begin{abstract}
    We derive boundary conditions at the interfaces of magnetoelastic heterostructures under ferromagnetic resonance for arbitrary magnetization directions and interface shapes. We apply our formalism to magnet$\vert$nonmagnet bilayers and magnetic grains embedded in a nonmagnetic thin film, revealing a nontrivial magnetization angle dependence of acoustic phonon pumping.
\end{abstract}

\maketitle
\end{CJK}

\section{Introduction}
The functionalities explored by spintronics may lead to novel information technologies that take advantage of the spin degrees of freedom. The current-induced spin transfer by electrons \cite{Sinova2015} and magnons \cite{Chumak2015,Brataas2020} can, e.g., be used for nonvolatile memories and magnetic logic devices. Electromagnetic fields \cite{Bliokh2015,Yang2020} and lattice vibrations \cite{Vonsovskii1962,Levine1962,McLellan1988,Niu2014,Kohno2018,Bliokh2019,Streib2020} may also carry spin. The orbital and spin angular momentum of the deformation fields of continuous isotropic acoustic media with SO(3) rotational symmetry derive from Noether's theorem \cite{Levine1962,Kohno2018}. The \textit{phonon spin} is the angular momentum contribution that does not depend on the origin of the coordinate system \cite{Kohno2018,Streib2020,Long2018}.

Magnetic anisotropy and magnetoelasticity in magnetic materials couple the phonon spin with the magnetization \cite{Kittel1958MEC}.
Interfaces between magnets (M) and nonmagnets (NM) play crucial roles in spintronics. An interesting material for the ``spin mechanics'' extension of spintronics are thin films of magnetic insulator yttrium iron garnet (YIG) grown on the paramagnet gadolinium gallium garnet (GGG), which are both of very high acoustic quality. The magnetization dynamics in YIG emits phonons into the GGG (\textit{phonon pumping}) \cite{Streib2018}. In YIG$\vert$GGG$\vert$YIG phononic spin valves, the magnetic layers communicate by the exchange of phonons over millimeters \cite{An2020,Andreas2020PRL}, much larger than the propagation distance of diffuse magnon spin currents in YIG \cite{Cornelissen2015}.

The Landau-Lifshitz-Gilbert (LLG) equation governs the magnetization dynamics and the elastic equation of motion (EOM) that of the underlying lattice. They are coupled by effective forces and fields, which are functional cross-derivatives of the total energy  \cite{Gurevich,Dreher2012,Kamra2015,Latcham2019,Godejohann2020}. This approach is appropriate in the GHz frequency regime in which wavelengths far exceed the lattice constants. Ferromagnetic resonance (FMR) excites the uniform precession (the Kittel mode) for which effective forces and torques in the bulk cancel out to a large extent. Dynamical magnetoelastic stresses at surfaces and interfaces of the magnet, however, are a source of phonons \cite{Comstock1963} and its generation is governed by boundary conditions (BCs).

Comstock and LeCraw \cite{Comstock1963} formulated the BCs for planar M$\vert$NM interfaces with magnetization normal to the plane. Tiersten \cite{Tiersten1964,Tiersten1965} addressed BCs of general structures such as sketched in Fig. \ref{fig_system} in the framework of nonlinear continuum mechanics, but the practical consequences of spin-lattice coupling are difficult to distill from the heavy mathematics. Here, we address the BCs in linear system, clearly separating elastic and magnetoelastic effect, generalizing Ref. \cite{Comstock1963} to arbitrary interface geometries and directions of the macrospin dynamics, including shear and pressure waves. We interpret the BCs in terms of physically appealing conservation laws for linear- and angular momentum currents. The formalism can handle different material combinations; here we focus on YIG$\vert$GGG and Galfenol$\vert$GaAs.
We illustrate the formalism by calculating the FMR spectra and phonon pumping for two geometries: M$\vert$NM bilayers and M grains embedded in NM thin film. The model leads to analytic expressions of FMR and phonon pumping in the planer systems, with a good agreement with the experiments by An \textit{et al}. \cite{An2020} when magnetization is normal to the interface, and we predict a magnetization angle-dependence that reveals generation of pressure waves. By the curvilinear BCs, the dynamics of magnetic grains, on the other hand, emits a nontrivial distribution of phonon spin currents.

This article is organized as follows.
In Sec. \ref{sec_formalism} we introduce and simplify Tiersten's formalism \cite{Tiersten1965} and define the linear- and angular-momentum phonon currents. Sec. \ref{sec_planar} and \ref{sec_disk} deal with phonon pumping from planar and curvilinear interfaces, respectively. In Sec. \ref{sec_discussion} we discuss applications and justify the approximations. In Sec. \ref{sec_conclusion} we summarize our results and give an outlook.

\section{\label{sec_formalism} Formalism}
\subsection{Variational Principle}
\begin{figure}
    \centering
    \includegraphics[width=8.6cm]{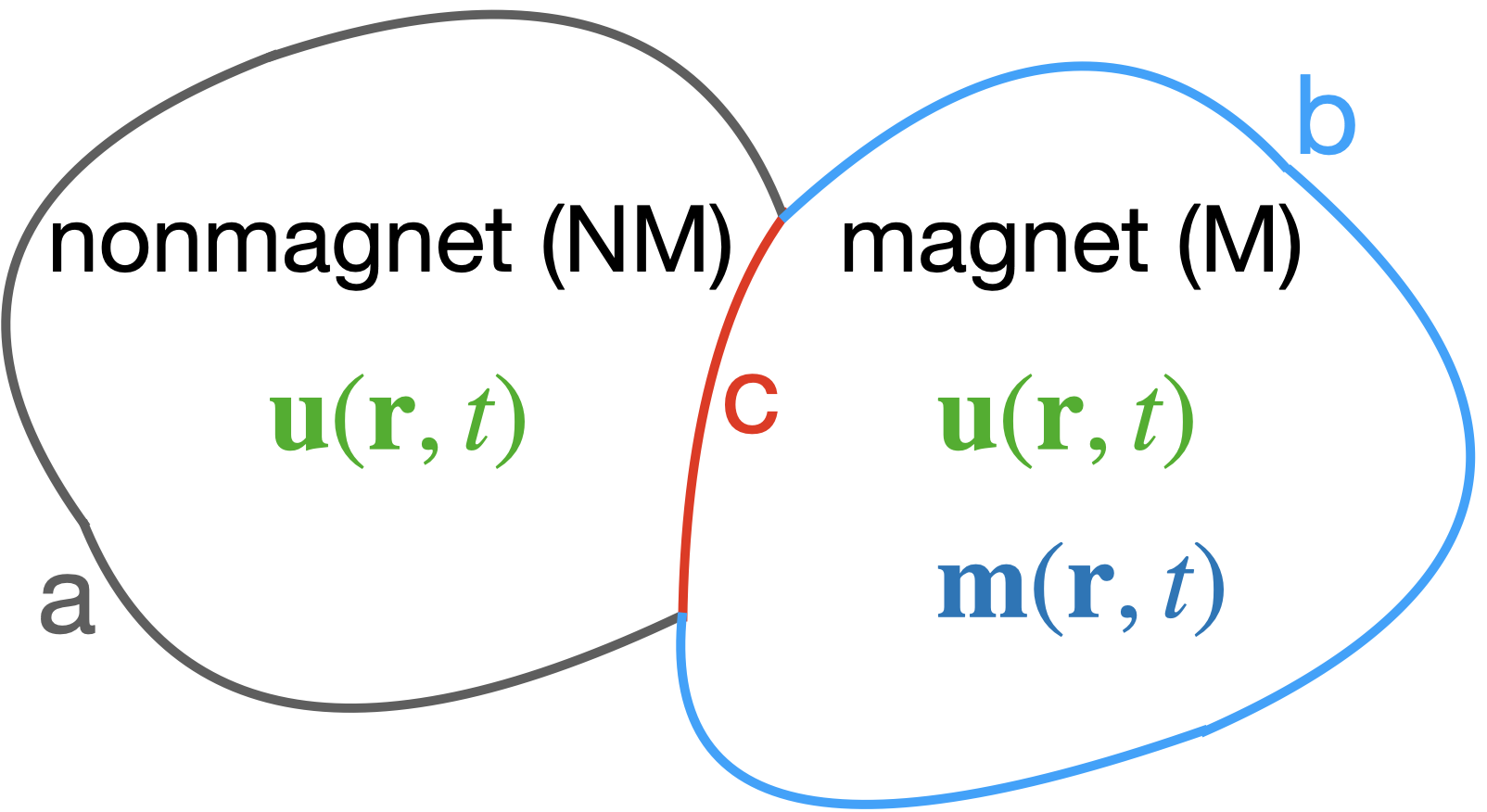}
    \caption{\label{fig_system} The displacement (deformation) vector fields of lattice $\mathbf{u}$ and magnetization $\mathbf{m}$ determine the state of magnet (M)-nonmagnet (NM) composite systems studied in this paper. The boundaries are (a) NM$\vert$vacuum surface (black), (b) M$\vert$vacuum surface (blue), and (c) NM$\vert$M interface (red).}
\end{figure}

We consider the M-NM composite system sketched in Fig. \ref{fig_system}. Sound waves in elastic media with frequencies up to tens of GHz have wavelengths much longer than the lattice constants and continuum theory applies. We disregard the effect of global rotations on magnons and phonons \cite{Andreas2020PRB}, assuming the total system size to be macroscopic. We focus on the linear regime in which the material (Lagrangian) and spatial (Eulerian) coordinates coincide \cite{Fung} and denote them by $\mathbf{r}$, while $\mathbf{u}(\mathbf{r},t)$ is the displacement vector at time $t$ of a volume element with equilibrium position $\mathbf{r}$. M and NM are assumed bonded, with identical displacement on both sides close to the interface (c) in Fig. \ref{fig_system}:
\begin{equation}\label{BC_ucontinuity}
    \mathbf{u}\vert_\mathrm{NM}=\mathbf{u}\vert_\mathrm{M} \quad\text{on (c)}.
\end{equation}
The magnetization vector field is normalized by its modulus $M_s$ as $\mathbf{m}(\mathbf{r},t)=\mathbf{M}(\mathbf{r},t)/M_s$.  Throughout, Greek letters $\alpha,\beta,\dots$ denote spatial coordinates and we adopt the summation convention over repeated indices.

We compartmentalize the Lagrangian density as
\begin{equation}\label{def_L}
    \mathcal{L}=
    \left\{
        \begin{aligned}
            &\mathcal{K}_\mathrm{el}-\mathcal{U}_\mathrm{el} &
            &\text{in }\mathrm{NM} \\
            &\mathcal{K}_\mathrm{el}-\mathcal{U}_\mathrm{el}+\mathcal{K}_\mathrm{mag}-\mathcal{U}_\mathrm{mag}-\mathcal{U}_\mathrm{surf}-\mathcal{U}_\mathrm{me}  &
            &\text{in }\mathrm{M}
        \end{aligned},
    \right.
\end{equation}
where the kinetic energy $\mathcal{K}_\mathrm{el}(\mathbf{r})$ is $\rho\dot{\mathbf{u}}(\mathbf{r})^2/2$ in NM and $\tilde{\rho}\dot{\mathbf{u}}(\mathbf{r})^2/2$ in M. The mass densities $\rho, \tilde{\rho}$ and constitutive parameters are taken to be constant in each material, but allowed to differ. The spin kinetic term reads $\mathcal{K}_\mathrm{mag}=M_s/(-\gamma)\dot{\phi}(\cos\theta-1)$, where $-\gamma$, $\theta$ and $\phi$ are the gyromagnetic ratio and the polar- and azimuthal angles, respectively \cite{BrownMicromagnetics,Nagaosa}. We consider elastic, magnetic, and magnetoelastic energy densities that depend linearly on the deformations, magnetization, and their derivatives
\begin{subequations}
    \begin{align}
        \mathcal{U}_\mathrm{el}&=\mathcal{U}_\mathrm{el}(\partial_\beta u_\alpha),  \label{def_U_el}\\
        \mathcal{U}_\mathrm{mag}&=\mathcal{U}_\mathrm{mag}(m_\alpha,\partial_\beta m_\alpha),  \label{def_U_mag}\\
        \mathcal{U}_\mathrm{me}&=\mathcal{U}_\mathrm{me}(\partial_\beta u_\alpha,m_\alpha,\partial_\beta m_\alpha). \label{def_U_me}
    \end{align}
\end{subequations}
The deformation gradients $\partial_\beta u_\alpha=\varepsilon_{\alpha\beta}+\omega_{\alpha\beta}$ consist of strains $\varepsilon_{\alpha\beta}=(\partial_\beta u_\alpha+\partial_\alpha u_\beta)/2$ and rotational deformations $\omega_{\alpha\beta}=(\partial_\beta u_\alpha-\partial_\alpha u_\beta)/2$. The coupling of strain and magnetization is the `magnetoelastic coupling' \cite{Kittel1958MEC,Gurevich}. Rotational deformations tilt the anisotropy axis, leading to the \textit{magnetorotation coupling} \cite{Maekawa1976,Garanin1997,Jaafar2009,Xu2020}. The general form (\ref{def_U_me}) includes both interactions but we still refer to it as \textit{magnetoelastic coupling} (MEC) in the following. The surface anisotropy energy $\mathcal{U}_\mathrm{surf}=\mu_0\mathbf{M}\cdot\mathbf{H}_\mathrm{surf}/2$ arises from crystalline and magnetodipolar effective fields $\mathbf{H}_\mathrm{surf}$ at the boundaries (b) and (c) \cite{Gurevich}.
The classical action functional reads
\begin{equation}
    S[\mathbf{u},\mathbf{m}]=\int_{0}^{T} dt\int_\mathrm{M+NM} d^3r\ \mathcal{L}(\dot{u}_\alpha,\partial_\beta u_\alpha,m_\alpha,\partial_\beta m_\alpha).
\end{equation}
We can derive the governing equations by the principle of least action. We assume a strong ferromagnet with constant modulus of the magnetization vector:
\begin{equation}\label{VarPrinciple}
    \delta S
    +\delta\int_0^T dt\left[\int_\mathrm{M} d^3r\ \lambda_1|\mathbf{m}|^2+\int_\mathrm{b+c}dS\ \lambda_2|\mathbf{m}|^2\right]
    =0.
\end{equation}
The Lagrange multipliers $\lambda_{1,2}$ enables independent variations of the three magnetization components $\{\delta m_\alpha\}$. In the absence of external forces at surfaces and interfaces, the variations at the boundaries must be taken into account when minimizing the action. For mathematical convenience, we impose $\delta u_\alpha(\mathbf{r},0)=\delta u_\alpha(\mathbf{r},T)=\delta m_\alpha(\mathbf{r},0)=\delta m_\alpha(\mathbf{r},T)=0$, which does not affect the results as long as the time $T$ is sufficiently large. Eq. (\ref{VarPrinciple}) yields the following EOMs and BCs for the lattice and magnetization \cite{Tiersten1965}:
\begin{subequations}\label{VarPrinciple_elEOM}
    \begin{align}
        \rho\ddot{u}_\alpha&=\frac{\partial\sigma_\mathrm{el}^{\alpha\beta}}{\partial r_\beta}& &\text{in }\mathrm{NM}, \label{VarPrinciple_elEOM_nm}\\
        \tilde{\rho}\ddot{u}_\alpha&=\frac{\partial\sigma_\mathrm{el}^{\alpha\beta}}{\partial r_\beta}+\frac{\partial\sigma_\mathrm{me}^{\alpha\beta}}{\partial r_\beta}& &\text{in }\mathrm{M}, \label{VarPrinciple_elEOM_m}
    \end{align}
\end{subequations}
\begin{subequations}\label{VarPrinciple_elBC}
    \begin{align}
        \bm{\sigma}_\mathrm{el}^\alpha\cdot\mathbf{n}&=0& &\text{on (a)}, \label{VarPrinciple_elBC_a}\\
        (\bm{\sigma}_\mathrm{el}^\alpha+\bm{\sigma}_\mathrm{me}^\alpha)\cdot\mathbf{n}&=0& &\text{on (b)}, \label{VarPrinciple_elBC_b}\\
        (\bm{\sigma}_\mathrm{el}^\alpha\vert_\mathrm{M}+\bm{\sigma}_\mathrm{me}^\alpha)\cdot\mathbf{n}&=\bm{\sigma}_\mathrm{el}^\alpha\vert_\mathrm{NM}\cdot\mathbf{n}& &\text{on (c)}, \label{VarPrinciple_elBC_c}
    \end{align}
\end{subequations}
\begin{equation}\label{VarPrinciple_LLeq}
    \dot{\mathbf{m}}=-\gamma\mu_0\mathbf{m}\times(\mathbf{H}_\mathrm{mag}+\mathbf{H}_\mathrm{me}) \quad \text{in }\mathrm{M},
\end{equation}
and
\begin{equation}\label{VarPrinciple_mBC}
    \epsilon_{\alpha\beta\gamma}m_\beta\left(\frac{1}{2}\mu_0M_sH_\mathrm{surf}^\gamma+\mathbf{X}^\gamma\cdot\mathbf{n}\right)=0
\end{equation}
on boundaries to M, where
\begin{subequations}
    \begin{align}
        \sigma_\mathrm{el}^{\alpha\beta}&=\frac{\partial\mathcal{U}_\mathrm{el}}{\partial(\partial_\beta u_\alpha)}, \quad
        \sigma_\mathrm{me}^{\alpha\beta}=\frac{\partial\mathcal{U}_\mathrm{me}}{\partial(\partial_\beta u_\alpha)}, \label{def_sigma_initial}\\
        \mathbf{H}_\mathrm{mag}&\equiv -\frac{1}{\mu_0M_s}\left[\frac{\partial\mathcal{U}_\mathrm{mag}}{\partial\mathbf{m}}-\partial_\nu\frac{\partial\mathcal{U}_\mathrm{mag}}{\partial(\partial_\nu\mathbf{m})}\right],\\
        \mathbf{H}_\mathrm{me}&\equiv -\frac{1}{\mu_0M_s}\left[\frac{\partial\mathcal{U}_\mathrm{me}}{\partial\mathbf{m}}-\partial_\nu\frac{\partial\mathcal{U}_\mathrm{me}}{\partial(\partial_\nu\mathbf{m})}\right],\\
        X^{\alpha\beta}&\equiv\frac{\partial\mathcal{U}_\mathrm{mag}}{\partial(\partial_\beta m_\alpha)}+\frac{\partial\mathcal{U}_\mathrm{me}}{\partial(\partial_\beta m_\alpha)},
    \end{align}
\end{subequations}
the vectors of tensor components are defined as $\mathbf{A}^\alpha\equiv(A^{\alpha x},A^{\alpha y},A^{\alpha z})^T$, and $\mathbf{n}$ is a surface normal. In Appendix \ref{app_stress} we discuss the definition of the stress tensor (\ref{def_sigma_initial}).

MEC causes a nonvanishing \textit{magnetoelastic stress} tensor $\sigma_\mathrm{me}$ and forces $F_\mathrm{me}^\alpha=\partial_\beta\sigma_\mathrm{me}^{\alpha\beta}$ in Eq. (\ref{VarPrinciple_elEOM_m}). A uniform magnetization appears to not affect the dynamics, but comes into play via the BCs \cite{Gurevich,Comstock1963}, as seen in Eqs. (\ref{VarPrinciple_elBC}) that ensures the continuity of stress across the interfaces and boundaries.
Eqs. (\ref{VarPrinciple_elBC}) contain all surface and interface stresses and uniquely determines the solution to Eqs. (\ref{VarPrinciple_elEOM}) \cite{Achenbach}.
In the absence of MEC ($\bm{\sigma}_\mathrm{me}^\alpha=0$), Eqs. (\ref{VarPrinciple_elBC}) reduce to the free- and bonded-boundary conditions in the ordinary theory of elasticity \cite{Graff}. Akhiezer \cite{Akhiezer} and Tiersten \cite{Tiersten1965} derived the BCs at the boundaries (b) and (c) more than half a century ago. However, these authors did not separate magnetoelastic and elastic contributions, which is helpful for practical implementations and physical understanding. Eqs. (\ref{VarPrinciple_elBC}) hold for arbitrarily curved interfaces, magnetization direction, and MEC energy.
For a bilayer system with macrospin magnetization normal or parallel to the plane, Eq. (\ref{VarPrinciple_elBC_c}) reduces to MEC-BCs involving only pure shear waves \cite{Comstock1963,Kamra2015,Streib2018,Andreas2020PRL}.

Magnetization dissipation can be taken into account in Eq. (\ref{VarPrinciple_LLeq}) by adding a phenomenological viscous damping torque to arrive at the LLG equation:
\begin{equation}\label{LLG_0}
    \dot{\mathbf{m}}=-\gamma\mu_0\mathbf{m}\times(\mathbf{H}_\mathrm{mag}+\mathbf{H}_\mathrm{me})+\alpha_\mathrm{G}\mathbf{m}\times\dot{\mathbf{m}},
\end{equation}
where $\alpha_\mathrm{G}$ is the Gilbert damping constant.
Ultrasonic attenuation in solids arises from thermoelasticity, phonon-phonon interactions, and defects and depends on frequency and temperature \cite{Truell}. Here we focus on a small frequency range and room temperature and model the attenuation by additional damping forces $-\rho\eta_\mathrm{el}\dot{u}_\alpha$ on the right-hand side of Eq. (\ref{VarPrinciple_elEOM}) with frequency-independent attenuation per unit length, for which we adopt the parameters from the MHz-GHz experiments (see Table \ref{table_materials}).
Both types of phenomenological damping, $\alpha_\mathrm{G}$ and $\eta_\mathrm{el}$, cause a loss of angular momentum that can in a microscopic description be accounted for by a transfer of angular momentum to global rotations \cite{Andreas2020PRB} or to the environment that supports the sample \cite{Streib2020}.

The magnetization obeys the BC (\ref{VarPrinciple_mBC}). In the long-wavelength limit the exchange contribution to the MEC vanishes, i.e., $\partial\mathcal{U}_\mathrm{me}/\partial(\partial_\beta m_\alpha)=0$. When, on the other hand, the surface anisotropy is small compared to exchange interaction, Eq. (\ref{VarPrinciple_mBC}) simplifies to `free' BC, i.e., vanishing magnetization gradient at the boundaries.

\subsection{\label{sec_jp} Linear momentum current}
According to Streib \textit{et al}. \cite{Streib2018} the BCs for the bilayer with perpendicular and in-plane magnetization reflect conservation of the linear momentum current at the interface. Here we extend this notion for arbitrary shape of the boundaries and magnetization directions. Newton's equations (\ref{VarPrinciple_elEOM}) are equivalent to the conservation law of linear momentum,
\begin{equation}\label{EoM_p}
    \frac{dp^\alpha}{dt}=-\mathrm{div}\mathbf{j}_p^\alpha,
\end{equation}
where
\begin{equation}\label{def_j}
    \mathbf{j}_p^\alpha(\mathbf{r}) \equiv
    \left\{
        \begin{aligned}
            &-\bm{\sigma}_\mathrm{el}^\alpha(\mathbf{r})& &\mathbf{r}\in\text{NM} \\
            &-\bm{\sigma}_\mathrm{el}^\alpha(\mathbf{r})
            -\bm{\sigma}_\mathrm{me}^\alpha(\mathbf{r}) & &\mathbf{r}\in\text{M}
        \end{aligned}
    \right.
\end{equation}
is the (outward) \textit{linear momentum current density tensor} with units $[\mathrm{N\ m}^{-2}]=[\mathrm{kg}\ \mathrm{m/s}\ \mathrm{s}^{-1}\ \mathrm{m}^{-2}]$ (linear momentum flux per unit area). The index $\alpha$ denotes Cartesian component of the linear momentum, whereas the vector is the current flow direction. The minus signs in Eq. (\ref{def_j}) indicate that the stress is a force exerted on the volume by its surrounding parts of the body \cite{Landauel} equivalent to an incoming flow of linear momentum. Our central result (\ref{VarPrinciple_elBC}) and assumption (\ref{BC_ucontinuity}) are therefore equivalent to the continuity of linear momentum current and displacement:
\begin{equation}\label{elBC_jp}
    \begin{aligned}
        \mathbf{j}_p^{\alpha}\cdot\mathbf{n}=0&& &\text{on outer surfaces (a),(b)}, \\
        \left.
            \begin{aligned}
                \mathbf{u}\vert_\mathrm{NM}&=\mathbf{u}\vert_\mathrm{M} \\
                \mathbf{j}_p^{\alpha}\vert_\mathrm{NM}\cdot\mathbf{n}&=\mathbf{j}_p^{\alpha}\vert_\mathrm{M}\cdot\mathbf{n}
            \end{aligned}
        \right\}&& &\text{on interface (c).}
    \end{aligned}
\end{equation}

\subsection{\label{sec_jS} Angular momentum currents}
Akhiezer \cite{Akhiezer} and Kamra \textit{et al}. \cite{Kamra2015} derived the BCs by considering energy flux and energy conservation (integral of motion) but did not address the angular momentum. Here, we find that the linear momentum current introduced in Sec. \ref{sec_jp} is closely related to the magnon-phonon angular momentum current across M$\vert$NM interfaces.

Let us consider a volume element of a deformed elastic magnet located at the vector sum of its equilibrium position and displacement, $\mathbf{r}+\mathbf{u}$. When $|\mathbf{u}|\ll|\mathbf{r}|$, the volume integral of physical quantities over a deformed body and equilibrium body with volume $V$ is the same. The motion of a volume element in continuous media may acquire Newtonian angular momentum $\mathbf{J}_\mathrm{ph}=\mathbf{L}_\mathrm{ph}+\mathbf{S}_\mathrm{ph}$, where \cite{Niu2014,Garanin2015,Kohno2018,Streib2020}
\begin{align}
    \mathbf{L}_\mathrm{ph}&=\int_{V}d^3r\ \mathbf{r}\times\mathbf{p},&
    \mathbf{S}_\mathrm{ph}&=\int_{V}d^3r\ \mathbf{u}\times\mathbf{p}.
\end{align}
The first integral expresses a global rotation of the body, depends on the choice of the origin, and vanishes for elastic plane waves with finite wavelengths \cite{Streib2020}. The second may be interpreted as  ``phonon spin'', which is caused by unidirectional rotations of mass particles around their equilibrium positions \cite{Garanin2015}.
$\mathbf{J}_\mathrm{ph}$ is generated not only by the magnetorotation coupling in the volume \cite{Garanin2015,Kohno2018}, but also by the angular momentum current through the boundaries, since according to Eq. (\ref{EoM_p}):
\begin{subequations}\label{LSdot}
    \begin{align}
        \dot{L}_\mathrm{ph}^\alpha
        &=-\int_{\partial V}\mathbf{j}_L^\alpha\cdot\mathbf{n}dS+\int_V T_L^\alpha d^3r, \label{Ldot}\\
        \dot{S}_\mathrm{ph}^\alpha
        &=-\int_{\partial V}\mathbf{j}_S^\alpha\cdot\mathbf{n}dS+\int_V T_S^\alpha d^3r, \label{Sdot}
    \end{align}
\end{subequations}
where $\partial V$ is the surface and
\begin{subequations}
    \begin{align}
        \mathbf{j}_L^\alpha&=\epsilon_{\alpha\beta\gamma}r_\beta\mathbf{j}_p^\gamma,&
        T_L^\alpha&=\epsilon_{\alpha\beta\gamma}j_p^{\gamma\beta}, \\
        \mathbf{j}_S^\alpha&=\epsilon_{\alpha\beta\gamma}u_\beta\mathbf{j}_p^\gamma,&
        T_S^\alpha&=\epsilon_{\alpha\beta\gamma}\partial_\nu u_\beta j_p^{\gamma\nu}.
    \end{align}
\end{subequations}
The angular momentum current densities $\mathbf{j}_{L,S}^\alpha$ are in units (angular momentum)/(area)/(time). The surface integrals in Eqs. (\ref{LSdot}) represent the angular momentum transfer across the boundaries that has not been considered in the bulk theory \cite{Garanin2015}. For levitating (non)magnets with stress-free boundary, the surface integrals vanishes owing to the BC (\ref{elBC_jp}). $T_{L,S}^\alpha$ are torque densities. $T_L^\alpha$ induces a rigid rotation of the body and therefore does not involve local deformations. On the other hand, $T_S^\alpha$ exerts local torques by elastic deformations and hence depends on the derivatives of the mechanical displacement. In NM, $T_L^\alpha$ vanishes due to the symmetry of the elastic stress tensor, while $T_S^\alpha$ may be disregarded since it is quadratic in the strain. In M, $T_L^\alpha$ is finite due to the anti-symmetric part of the magnetoelastic stress tensor [see Eq. (\ref{def_sigma_matrix}) in Appendix \ref{app_stress}], which actuates a rigid rotation \cite{Garanin2015}. In the examples illustrated in Secs. \ref{sec_planar} and \ref{sec_disk}, however, the anti-symmetric part (i.e., magnetorotation coupling) is negligible and angular momenta are mostly supplied by currents across the boundaries. In Sec. \ref{sec_disk} we calculate the phonon spin current density $\mathbf{j}_S^\alpha$ emitted from a magnetic disk.

\section{\label{sec_planar} Applications: Bilayers}
As a first example, we consider a flat interface between a magnetic film of thickness $d$ and a nonmagnetic substrate of thickness $L$ [Fig. \ref{fig_AngleDependence}(a)]. The governing equations derived in Sec. \ref{sec_formalism} allows us to compute the FMR signals as a function of layer thicknesses and magnetization orientation, thereby microscopically modeling the published experiments with perpendicular magnetization \cite{An2020}.

\subsection{Model}
We take the $z$ axis normal to the interface and assume translational symmetry in the $x-y$ plane. We consider cubic lattices, whose elastic energy is given by three elastic stiffness constants $C_{ij}$ and the strains \cite{Kittel_SolidState}:
\begin{align}
    \mathcal{U}_\mathrm{el}=&\frac{C_{11}}{2}(\varepsilon_{xx}^2+\varepsilon_{yy}^2+\varepsilon_{zz}^2) \notag\\
    &+C_{12}(\varepsilon_{yy}\varepsilon_{zz}+\varepsilon_{zz}\varepsilon_{xx}+\varepsilon_{xx}\varepsilon_{yy}) \notag\\
    &+2C_{44}(\varepsilon_{yz}^2+\varepsilon_{zx}^2+\varepsilon_{xy}^2). \label{U_el}
\end{align}
Substitution into the first of Eq. (\ref{def_sigma_initial}) reproduces Hooke's law.
In YIG and GGG, $C_{11}-C_{12}\simeq2C_{44}$ and the sound velocities are virtually isotropic \cite{Gurevich}, but in general anisotropic in single cubic crystals. We adopt the crystallographic orientation $\mathbf{e}_z\parallel \langle100\rangle$, so the transverse and longitudinal velocities of the ultrasounds propagating in the $z$ direction read $c_t=\sqrt{C_{44}/\rho}$ and $c_\ell=\sqrt{C_{11}/\rho}$, respectively.

The magnetic energy density consists of the Zeeman coupling $\mathcal{U}_Z=-\mu_0M_s\mathbf{m}\cdot\mathbf{H}_\mathrm{ext}$ and the shape and crystalline anisotropy energies \cite{Kittel1949}
\begin{align}
    \mathcal{U}_A&=\frac{1}{2}\mu_0M_s^2\mathbf{m}^T\mathsf{N}\mathbf{m}+K_1\left(\mathbf{m}\times\mathbf{n}_0\right)^2 \notag\\
    &=K_1+\left(\frac{1}{2}\mu_0M_s^2-K_1\right)(\mathbf{m}\cdot\hat{\mathbf{e}}_z)^2, \label{def_UA}
\end{align}
where $\mu_0$ is the vacuum permeability, $\mathsf{N}=\mathrm{diag}(0,0,1)$ is the demagnetization tensor of thin films, and $K_1$ the uniaxial crystalline anisotropy constant. Here we adopt a perpendicular anisotropy axis $\mathbf{n}_0\parallel\hat{\mathbf{e}}_z$. We disregard surface anisotropies which could pin the magnetization at the boundaries \cite{Kittel1958SW,Guslienko2005,Wang2019}. The uniform equilibrium magnetization is assumed to be parallel to a strong enough applied magnetic field $\mathbf{H}_0$. The Kittel mode is excited by a weak AC magnetic field transverse to the magnetization.

The equilibrium magnetization lies in the $z-x$ plane [Fig. \ref{fig_AngleDependence}(a)] at an angle $\theta_m$ with the $z$ axis. The external field and magnetization consist of static and dynamical components
\begin{equation}
    \mathbf{H}_\mathrm{ext}(t)
    =\mathcal{R}_y(\theta_m)\begin{pmatrix}
        h_\parallel(t) \\ h_\perp(t) \\ H_0
    \end{pmatrix},
\end{equation}
\begin{equation}\label{generalm}
    \mathbf{m}(\mathbf{r},t)\simeq\mathbf{m}(t)
    \simeq\mathcal{R}_y(\theta_m)\begin{pmatrix}
        m_\parallel(t) \\ m_\perp(t) \\ 1
    \end{pmatrix},
\end{equation}
where
\begin{equation}
    \mathcal{R}_y=
    \begin{pmatrix}
        \cos\theta_m & 0 & \sin\theta_m \\
        0 & 1 & 0 \\
        -\sin\theta_m & 0 & \cos\theta_m
    \end{pmatrix}
\end{equation}
is the rotation matrix around the $y$ axis. In Eq. (\ref{generalm}), we have assumed homogeneous and small magnetization amplitudes $m_{\parallel,\perp}\ll1$.

We expand the continuum MEC energy to linear order in strain and rotation tensor elements,
\begin{equation}\label{U_me}
    \mathcal{U}_\mathrm{me}
    =m_\alpha m_\beta\left[b_{\alpha\beta}\varepsilon_{\alpha\beta}+K_{\alpha\beta}\omega_{\alpha\beta}\right],
\end{equation}
where $b_{\alpha\beta}=\delta_{\alpha\beta}b_1+(1-\delta_{\alpha\beta})b_2$ and $b_{1,2}$ are the MEC parameters. The magnetorotation coupling arises from the rotation of the hard anisotropy axis $\hat{\mathbf{e}}_z\to\hat{\mathbf{e}}_z+\delta\mathbf{n}$, where $\delta\mathbf{n}=(\nabla\times\mathbf{u})\times\hat{\mathbf{e}}_z/2=(\omega_{xz},\omega_{yz},0)$ \cite{Jaafar2009,Streib2018}. From Eq. (\ref{def_UA}) we derive $K_{xz}=K_{yz}=-K_1+\mu_0M_s^2/2$, $K_{zx}=K_{zy}=K_1-\mu_0M_s^2/2$, while the other components vanish.

\subsection{Magnetization dynamics}
We derive from Eqs. (\ref{def_UA})(\ref{U_me}) the anisotropy and effective fields:
\begin{align}
    \gamma\mu_0\mathbf{H}_A
    &=-\frac{\gamma}{M_s}\frac{\partial\mathcal{U}_A}{\partial\mathbf{m}}
    =-(\omega_M-\omega_K)m_z\mathbf{e}_z, \label{Heff_A} \\
    \gamma\mu_0\mathbf{H}_\mathrm{me}
    &=-\frac{\gamma}{M_s}\frac{\partial\mathcal{U}_\mathrm{me}}{\partial\mathbf{m}}
    =-\mathsf{\Omega}_\mathrm{me}\mathcal{R}_y
    \begin{pmatrix}
        m_\parallel \\ m_\perp \\ 1
    \end{pmatrix}, \label{Heff_me}
\end{align}
where $\omega_K=\gamma 2K_1/M_s$, $\omega_M=\gamma\mu_0M_s$ and $\mathsf{\Omega}_\mathrm{me}^{\alpha\beta}=\frac{2\gamma}{M_s}[b_{\alpha\beta}\varepsilon_{\alpha\beta}+K_{\alpha\beta}\omega_{\alpha\beta}]$ are angular frequencies. From Eqs. (\ref{LLG_0})(\ref{generalm})(\ref{Heff_A})(\ref{Heff_me}) we obtain the linearized LLG equation in frequency domain:
\begin{equation}\label{LLG_1}
    \begin{pmatrix}
        m_\parallel \\ m_\perp
    \end{pmatrix}
    (\omega)
    =\mathsf{\chi}_\mathrm{FMR}(\omega,\theta_m)
    \left[
        \begin{pmatrix}
            h_\parallel \\ h_\perp
        \end{pmatrix}
        -\frac{1}{\gamma\mu_0}
        \begin{pmatrix}
            \Omega_\mathrm{me}^{\prime 13} \\ \Omega_\mathrm{me}^{\prime 23}
        \end{pmatrix}
    \right](\omega),
\end{equation}
where $\mathsf{\Omega}_\mathrm{me}^\prime =\mathcal{R}_y^{-1}\mathsf{\Omega}_\mathrm{me}\mathcal{R}_y$ represents the MEC, whereas the susceptibility tensor
\begin{equation}\label{pureFMR}
    \mathsf{\chi}_\mathrm{FMR}(\omega,\theta_m)=
    \frac{\gamma\mu_0}{\Delta_\mathrm{FMR}}
    \begin{pmatrix}
        \omega_{11}^0 & -i\omega \\
        i\omega & \omega_{22}^0
    \end{pmatrix}
\end{equation}
governs the pure FMR. The determinant $\Delta_\mathrm{FMR}=\omega_{11}^0\omega_{22}^0-\omega^2$ and the matrix elements $\omega_{11}^0=\omega_H-(\omega_M-\omega_K)\cos^2{\theta_m}-i\alpha_\mathrm{G}\omega$, $\omega_{22}^0=\omega_H-(\omega_M-\omega_K)\cos{2\theta_m}-i\alpha_\mathrm{G}\omega$. For $\theta_m\neq0^\circ$ the magnetization precession is elliptic. The \textquotedblleft tickle \textquotedblright field \cite{Weiler2011} in Eq. (\ref{LLG_1}) is induced by the lattice strains and rotations. For the present case $\mathbf{u}(\mathbf{r},\omega)=\mathbf{u}(z,\omega)$ and
\begin{equation}\label{Omega_epsilon}
    \begin{aligned}
        \Omega_\mathrm{me}^{\prime 13}&=\omega_c\partial_zu_x \cos{2\theta_m}-\omega_c^\ell \partial_zu_z\sin{2\theta_m}, \\
        \Omega_\mathrm{me}^{\prime 23}&=\omega_c\partial_zu_y \cos{\theta_m},
    \end{aligned}
\end{equation}
where $\omega_c=\omega_M/2+\gamma (b_2-K_1)/M_s$ and $\omega_c^\ell =\gamma b_1/M_s$ parameterize the magnetostriction and magnetorotation coupling.

\subsection{\label{sec_MEstress} Magnetoelastic surface stresses}

\begin{figure}
    \centering
    \includegraphics[width=8.6cm]{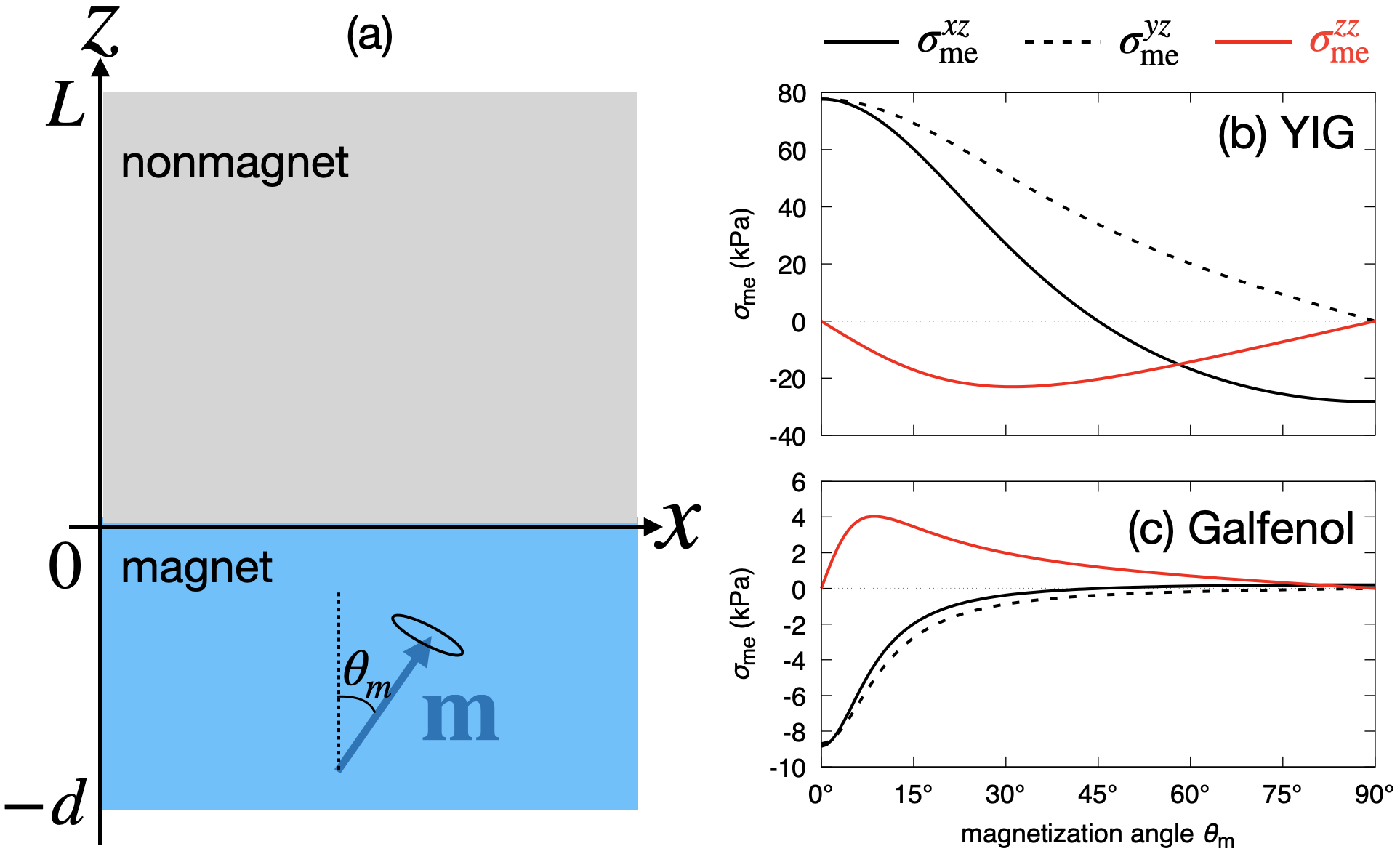}
    \caption{\label{fig_AngleDependence} (a) Sketch of a planar M$\vert$NM bilayer with magnetization angle $\theta_m$ and film thicknesses $L$ and $d$. (b,c) $\theta_m$-dependence of the amplitudes of the dynamical magnetoelastic interface stress (\ref{sigma_me_1D_om}) induced by pure FMR for different magnets with material parameters from Table \ref{table_materials}. Magnetization aligns the applied static fields of (b) $\mu_0H_0=0.3609$ T and (c) 1.6 T and is excited with the same perpendicular microwave intensity.}
\end{figure}

\begin{table*}
    \caption{\label{table_materials} Material parameters used in this paper.}
    \begin{ruledtabular}
        \begin{tabular}{llllllllll}
            & $\rho\ (\mathrm{kg/m^3})$ & $c_t\ (\mathrm{m/s})$ & $c_\ell\ (\mathrm{m/s})$ & $\eta_\mathrm{el}/(2\pi)$ (MHz) & $\alpha_\mathrm{G}$ & $\mu_0M_s$ (T) & $K_1\ (\mathrm{MJ/m^3})$ & $b_1\ (\mathrm{MJ/m^3})$ & $b_2\ (\mathrm{MJ/m^3})$ \\\hline
            YIG & 5170 \footnotemark[1] & 3843 \footnotemark[1] & 7209 \footnotemark[1] & 0.35 & 9$\times10^{-5}$ \footnotemark[2] & 0.172 \footnotemark[2] & -6.10$\times10^{-4}$ \footnotemark[3] & 0.348 \footnotemark[1] & 0.696 \footnotemark[1] \\
            GGG & 7070 \footnotemark[4] & 3568 \footnotemark[4] & 6411 \footnotemark[4] & 0.35 \footnotemark[2] &  &  &  &  &  \\
            Galfenol & 7800 \footnotemark[5] & 4.0$\times10^3$ \footnotemark[5] & 5.0$\times10^3$ \footnotemark[6] & 1 & 0.017 \footnotemark[7] & 1.59 \footnotemark[5] & 3.3$\times10^{-2}$ \footnotemark[7] & -15.8 \footnotemark[7] & -6.19 \footnotemark[5] \\
            GaAs & 5317 \footnotemark[9]& 3.34$\times10^3$ \footnotemark[9] & 4.73$\times10^3$ \footnotemark[9] & 1 \footnotemark[10] &  &  &  &  &
        \end{tabular}
        \footnotetext[1]{Ref. \cite{Gurevich}.}
        \footnotetext[2]{Ref. \cite{An2020}.}
        \footnotetext[3]{Ref. \cite{Hansen1974}.}
        \footnotetext[4]{Ref. \cite{Kleszczewski1988}.}
        \footnotetext[5]{Ref. \cite{Godejohann2020}.}

        \footnotetext[6]{Calculated from $C_{44}=1.23\times10^{11}$ Pa and $C_{11}=1.96\times10^{11}$ Pa \cite{Kellogg2004}.}
        \footnotetext[7]{Thin film value \cite{Parkes2013}.}
        \footnotetext[8]{Bulk single crystal value \cite{Clark2003}.}
        \footnotetext[9]{Ref. \cite{Blakemore1982}. For propagation along $\langle 100\rangle$ and at 300 K.}
        \footnotetext[10]{In the absence of experimental data at 3-10 GHz, we average the room-temperature results at 1.03 GHz \cite{Helme1978} [$\eta_\mathrm{el}/(2\pi)\sim0.1$ MHz] and 56 GHz \cite{Chen1994} ($\sim60$ MHz).}
    \end{ruledtabular}
\end{table*}

Magnetization precession at frequency $\omega$ induces ac surface stresses on the $x-y$ plane. In frequency space, Eq. (\ref{def_sigma_initial}) reduces to
\begin{equation}\label{sigma_me_1D_om}
    \begin{pmatrix}
        \sigma_\mathrm{me}^{xz} \\ \sigma_\mathrm{me}^{yz} \\ \sigma_\mathrm{me}^{zz}
    \end{pmatrix}
    (\omega)
    =
    \begin{pmatrix}
        \left(b_2-K_1+\frac{1}{2}\mu_0M_s^2\right)m_\parallel(\omega)\cos{2\theta_m} \\
        \left(b_2-K_1+\frac{1}{2}\mu_0M_s^2\right)m_\perp(\omega)\cos{\theta_m} \\
        -b_1 m_\parallel(\omega)\sin{2\theta_m}
    \end{pmatrix},
\end{equation}
where we discarded static ($\omega=0$) as well as higher order terms in the transverse magnetization, implying that the stress (not the strain) adiabatically follows the magnetization precession \cite{Rinaldi1985}. For YIG, $K_1/b_2=0.0009$, $(\mu_0M_s^2/2)/b_2=0.02$, so $b_2$ dominates [Eq. (\ref{sigma_me_1D_om})]. In iron gallium alloy ($\mathrm{Fe_{0.81}Ga_{0.19}}$, Galfenol), $K_1/b_2=0.005$, $(\mu_0M_s^2/2)/b_2=0.1$, so the magnetorotation coupling due to dipolar anisotropy may become significant.

The angular dependencies of the magnetoelastic stresses (\ref{sigma_me_1D_om}) in YIG and Galfenol are plotted in Fig. \ref{fig_AngleDependence}(b) and (c), respectively, for the Kittel mode excited by the pure FMR (\ref{pureFMR}). The shear stresses $\sigma_\mathrm{me}^{xz}$ and $\sigma_\mathrm{me}^{yz}$ are maximal for $\theta_m=0^{\circ}$. While the former remains finite at $\theta_m=90^{\circ}$, the latter vanishes, leading to less efficient pumping of linearly polarized phonons \cite{Streib2018}.
The pressure force vanishes at $\theta_m=0^{\circ}$ and $90^{\circ}$, but at intermediate angles pumps longitudinal phonons, as discussed in the next subsection. For fixed microwave intensity, the Kittel mode amplitude in Galfenol is smaller due to the large Gilbert damping, which reduces the magnetoelastic stresses despite its large MEC parameters.

\subsection{\label{subsec_1Dplanar} 1D phonon pumping}
The phonon pumping problem derived in Sec. \ref{sec_formalism} can be solved analytically by a plane-wave ansatz. We first solve the elastic EOM and substitute it into the BCs to find the relation between the elastic wave amplitudes and magnetization. The strain effective field in Eq. (\ref{LLG_1}) is then proportional to the magnetization and affects the magnetic response. The imaginary part of the magnetic susceptibility is proportional to the microwave power absorption.

The EOM (\ref{EoM_p}) in M for sound waves propagating normal to the film separates into three modes, two degenerate transverse ($\alpha=x,y$) and a longitudinal, which solve
\begin{subequations}\label{1D_EOM}
    \begin{align}
        \frac{\partial^2 u_\alpha}{\partial z^2}+\frac{\omega^2}{\tilde{c}_t^2}\left(1+i\frac{\tilde{\eta}_\mathrm{el}}{\omega}\right)u_\alpha&=0, \\
        \frac{\partial^2 u_z}{\partial z^2}+\frac{\omega^2}{\tilde{c}_\ell^2}\left(1+i\frac{\tilde{\eta}_\mathrm{el}}{\omega}\right)u_z&=0.
    \end{align}
\end{subequations}
Similar equations without tilde hold in NM. The sound velocities and phenomenological ultrasonic attenuation parameters are summarized in Table \ref{table_materials}, where we assume the same attenuation parameter for the magnetic film and the substrate. The characteristic attenuation length of the TA modes is $\delta=c_t/\eta_\mathrm{el}=1.7$ mm in GGG and 0.5 mm in GaAs. The general solution to Eq. (\ref{1D_EOM}) is
\begin{align}
    u_{\alpha}(z,\omega)&=\tilde{A}^\alpha e^{i\tilde{k}_t^\prime z}+\tilde{B}^\alpha e^{-i\tilde{k}_t^\prime z}, \label{1D_Ansatz}\\
    \tilde{k}_t^\prime &=\frac{\omega}{\tilde{c}_t}\sqrt{1+i\frac{\eta_\mathrm{el}}{\omega}}
    \approx \tilde{k}_t+i\tilde{\kappa}_t, \notag
\end{align}
where $\tilde{k}_t=\omega/\tilde{c}_t$ and the damping parameter $\tilde{\kappa}_t=\tilde{\eta}_\mathrm{el}/2\tilde{c}_t$. The BCs (\ref{elBC_jp}) read for $\alpha=x,y,z$,
\begin{equation}\label{1D_BCs}
    \left\{
        \begin{aligned}
            u_\alpha(0^-)&=u_\alpha(0^+) \\
            \sigma_\mathrm{el}^{\alpha z}(-d)+\sigma_\mathrm{me}^{\alpha z}(-d)&=0 \\
            \sigma_\mathrm{el}^{\alpha z}(0^-)+\sigma_\mathrm{me}^{\alpha z}(0^-)&=\sigma_\mathrm{el}^{\alpha z}(0^+) \\
            \sigma_\mathrm{el}^{\alpha z}(L)&=0
        \end{aligned}
    \right. ,
\end{equation}
which should be used with Eq. (\ref{sigma_me_1D_om}). Since the Kittel mode feels only the spatial average of the effective field, we average the strain in Eq. (\ref{Omega_epsilon}) over the film thickness and rewrite it in terms of transverse magnetization using Eq. (\ref{1D_BCs}) (see Appendix \ref{app_1D}) such that
\begin{equation}\label{LLG_renorm}
    \begin{pmatrix}
        m_\parallel \\ m_\perp
    \end{pmatrix}
    (\omega)
    =\mathsf{\chi}_\mathrm{tot}(\omega,\theta_m)
    \begin{pmatrix}
        h_\parallel \\ h_\perp
    \end{pmatrix}(\omega).
\end{equation}
The susceptibility tensor
\begin{equation}\label{chi_tot}
    \mathsf{\chi}_\mathrm{tot}(\omega,\theta_m)
    =\frac{\gamma\mu_0}{\Delta(\omega,\theta_m)}
    \begin{pmatrix}
        \omega_{11} & -i\omega \\
        i\omega & \omega_{22}
    \end{pmatrix},
\end{equation}
where $\Delta(\omega,\theta_m)=\omega_{11}\omega_{22}-\omega^2$, includes the coupling to the lattice. The matrix elements $\omega_{11}=\omega_{11}^0-g(\omega)\cos^2{\theta_m}$ and $\omega_{22}=\omega_{22}^0-g(\omega)\cos^2{2\theta_m}-g^\ell (\omega)\sin^2{2\theta_m}$ are shifted by the complex coupling strengths (in units of angular frequency)
\begin{subequations}\label{def_coupling}
    \begin{align}
        g(\omega)
        &=\frac{M_s}{\gamma d\tilde{\rho}\tilde{c}_t}\frac{\omega_c^2}{\omega+i\tilde{\eta}_\mathrm{el}/2}F(\omega), \label{def_g}\\
        g^\ell (\omega)
        &=\frac{M_s}{\gamma d\tilde{\rho}\tilde{c}_\ell}\frac{(\omega_c^\ell )^2}{\omega+i\tilde{\eta}_\mathrm{el}/2}F^\ell(\omega), \label{def_gl}
    \end{align}
\end{subequations}
and depend on magnetization orientation. The real and imaginary part of Eq. (\ref{def_coupling}) modify the anisotropy fields and damping torques, respectively.
$F(\omega)$ and $F^\ell(\omega)$ are complex function of system geometry and material parameters (Appendix \ref{app_1D}). The coupling strengths in our microscopic theory includes the effect of acoustic damping and scales as $\sim\omega^{-1}$, while in the simple coupled oscillator models \cite{An2020,Litvinenko2020} the coupling is independent of the damping and tends to $\sim\omega^{-1/2}$. These differences may be important when a wider frequency range is of interest and the frequency dependence of $\tilde{\eta}_\mathrm{el}$ becomes significant \cite{Truell,David}. For the out-of-plane ($\theta_m=0^\circ$) configuration $\chi_\mathrm{tot}^{11}=\chi_\mathrm{tot}^{22}$ induces circular precession, whereas for other angles the precession is elliptic. The microwave power absorption
\begin{equation}\label{Pabs}
    P_\mathrm{abs}(\omega,\theta_m)\propto\mathrm{Im}(\mathbf{h}^T\mathbf{m})=\mathrm{Im}\chi_\mathrm{tot}^{11}h_\parallel^2+\mathrm{Im}\chi_\mathrm{tot}^{22}h_\perp^2
\end{equation}
is the observable in FMR experiments.

\begin{figure*}
    \includegraphics[width=17.2cm]{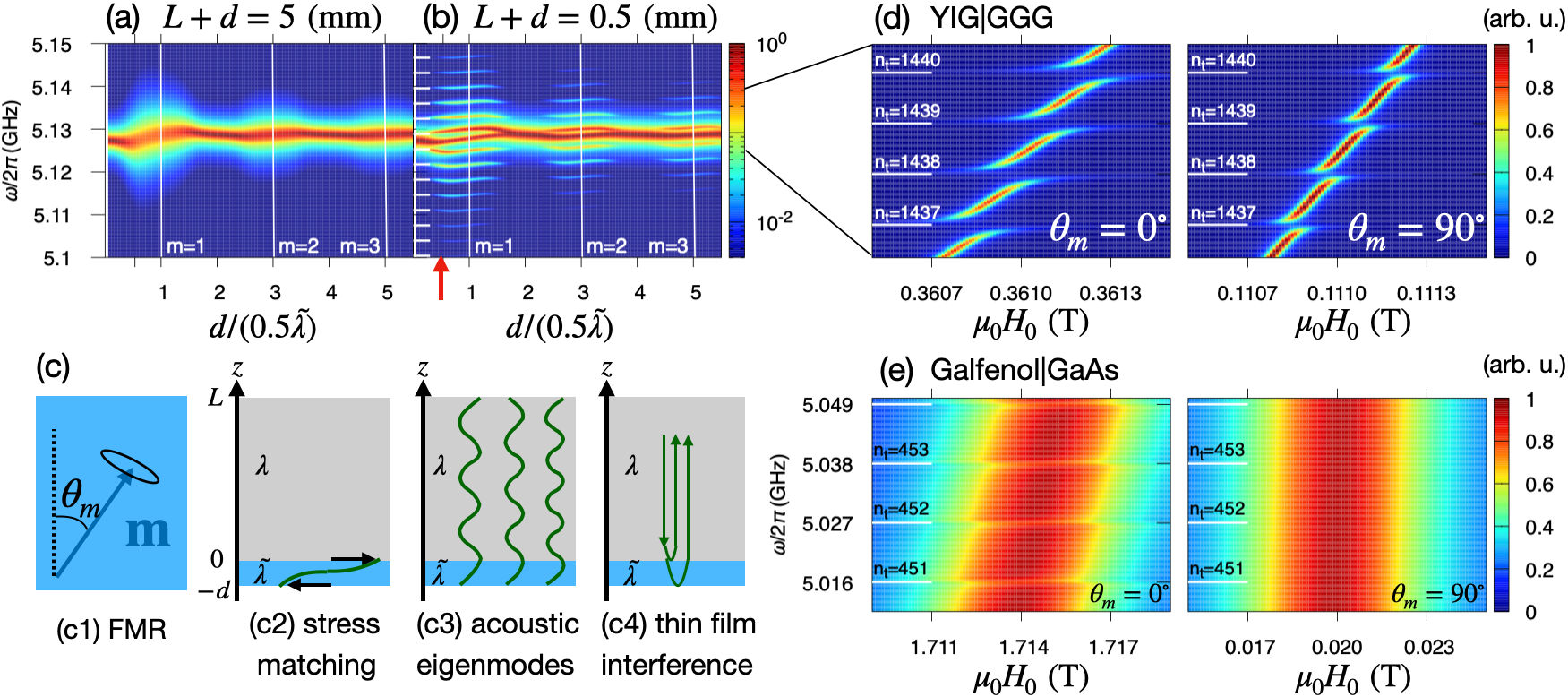}
    \caption{\label{fig_MHzAnticrossing} Phonon pumping in a YIG$\vert$GGG bilayer. (a,b) FMR absorption spectrum [Eq. (\ref{Pabs})] under a perpendicular static field $\mu_0H_0=0.3609$ (T) that pulls the magnetization fully out of plane ($\theta_m=0^\circ$) as a function of the frequency of the microwave and magnetic film thickness $d$ for bilayer thickness of (a) 5 mm and (b) 0.5 mm, normalized by half of the TA phonon wavelength at the FMR frequency. The red arrow indicates the film thickness $d/(0.5\tilde{\lambda})=0.53$ in the experiments \cite{An2020}. White lines labeled by the mode index $m$ indicate the stress-matching conditions [Eq. (\ref{StressMatching})]. The ladder on the ordinate in (b) marks the eigenfrequencies [Eq. (\ref{EigenFreqt})] of the TA standing waves with mode numbers from $n_t=1430$ (bottom) to 1443 (top) [see also (d) and (e)].
    (c) Schematic of the resonances in M$\vert$NM bilayer. Green curves represent acoustic waves. (d) shows the absorption fine structure in the vicinity of the FMR frequency as a function of applied static field for normal to and in-plane magnetization, respectively ($d=200$ nm, $L=0.5$ mm). (e) The corresponding plots for Galfenol$\vert$GaAs bilayers. The contour color scales are normalized by the maximum values.}
\end{figure*}

The absorption spectrum (\ref{Pabs}) contains the resonances sketched in Fig. \ref{fig_MHzAnticrossing}. Panel (c1) illustrates the FMR frequency dependence on magnetic anisotropies, external field, and magnetization orientation.
The resonance (c2) occurs when the magnetoelastic surface stresses acting in the opposite directions on the two magnetic surfaces excite odd acoustic waves, which requires that the magnetic film thickness fulfills the \textit{stress-matching condition}
\begin{equation}\label{StressMatching}
    \tilde{k}_td=\pi(2m-1)
    \qquad(m=1,2,\dots).
\end{equation}
Under this condition the lattice displacement and the additional FMR broadening are maximized \cite{Streib2018,Andreas2020PRL}. These resonances are very broad for YIG$\vert$GGG because of the strong coupling at the interface.
(c3) Standing sound waves form when
\begin{equation}\label{EigenFreq_eq}
    \sin{\tilde{k}_td}\cos{k_tL}+\frac{\rho c_t}{\tilde{\rho}\tilde{c}_t}\cos{\tilde{k}_td\sin{k_tL}}=0.
\end{equation}
The acoustic impedance mismatch $\rho c_t/(\tilde{\rho}\tilde{c}_t)=1.27$ between YIG and GGG or $0.87$ between Galfenol and GaAs (Table \ref{table_materials}) is not important at GHz frequencies. The acoustic resonance frequencies then simplify to
\begin{equation}\label{EigenFreqt}
    f_{n_t}=\frac{n_t}{2\left(d/\tilde{c}_t+L/c_t\right)}\qquad(n_t=1,2,\dots).
\end{equation}
The same equation holds for the pressure waves by replacing the transverse by the longitudinal sound velocities. When the film thickness exceeds the sound attenuation length $\delta$ the back and forth reflected waves cannot interfere anymore and the discrete spectrum is smeared out into a continuous one, which is the regime considered by Streib \textit{et al}. \cite{Streib2018}.
The phase matching of the acoustic waves reflected by the two boundaries $z=-d$ and $z=0$ may also enhance the phonon pumping:
\begin{equation}\label{ThinFilm}
    2\tilde{k}_td=\pi\left(2s-1\right)
    \qquad(s=1,2,\dots),
\end{equation}
which we call (c4) \textit{thin-film interference} condition (Fig. \ref{fig_MHzAnticrossing}).

\begin{figure*}
    \includegraphics[width=17.2cm]{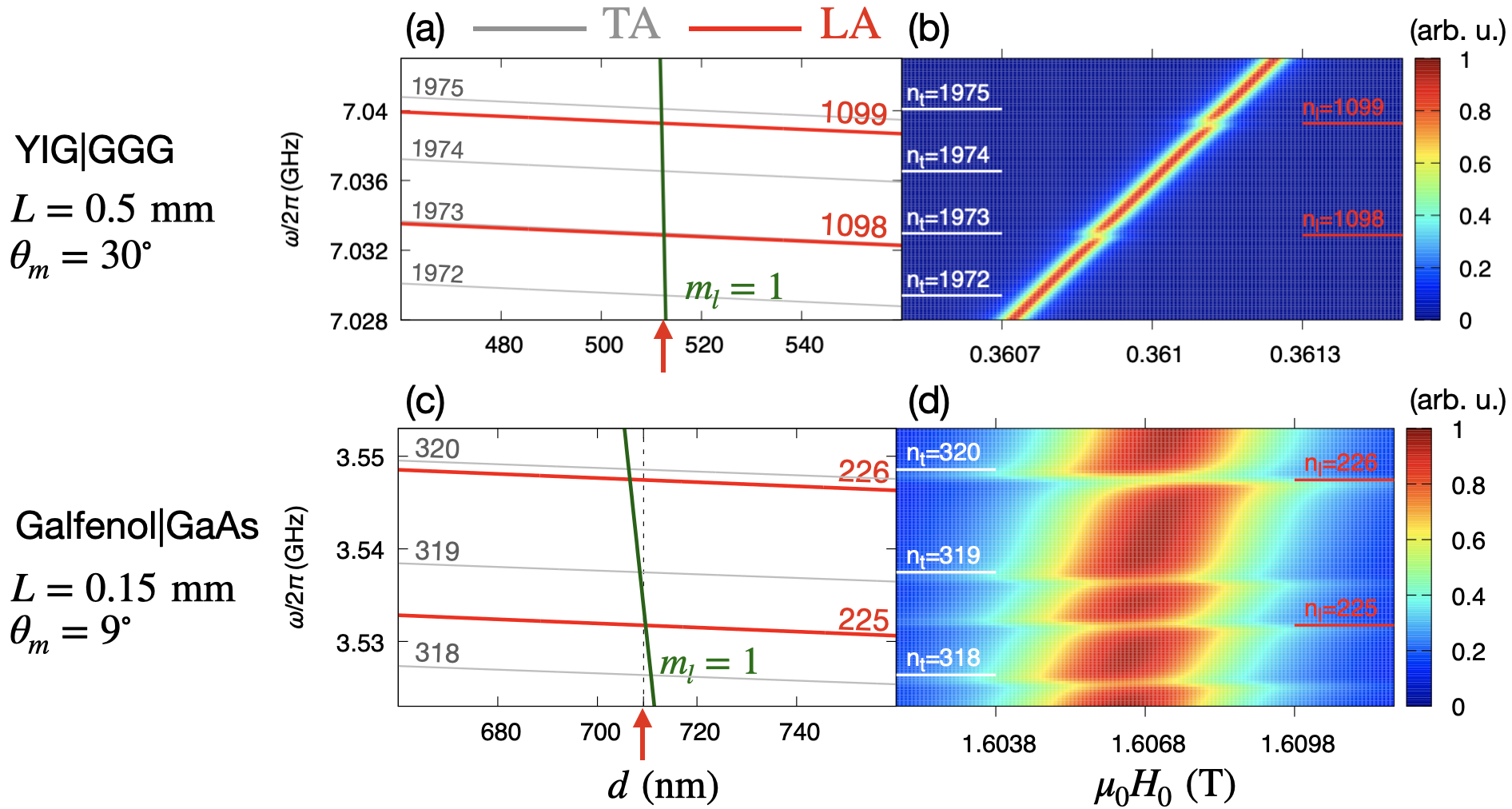}
    \caption{\label{fig_LAAnticrossing} Coupling to longitudinal phonons in different bilayers. (a,c) Acoustic eigenfrequencies (TA in gray and LA in red) and LA stress-matching condition (green) against the magnetic film thickness. TA stress-matching is not achieved in the plotted parameter space. FMR spectra on the right panels are for the film thicknesses indicated by the red arrows on the left panels and normalized (b) by the same factor as in Fig. \ref{fig_MHzAnticrossing} and (d) by the maximum value.}
\end{figure*}

We first focus on the normal $(\theta_m=0^\circ)$ configuration, in which the FMR excites only the transverse acoustic (TA) modes (see Fig. \ref{fig_AngleDependence}). In Fig. \ref{fig_MHzAnticrossing}(a) we plot the FMR spectrum [Eq. (\ref{Pabs})] of a YIG$\vert$GGG bilayer system with large thickness $L+d=5\  (\mathrm{mm})>\delta/2$. The external field of $0.3609$ T leads to the $f_\mathrm{FMR}=5.129$ GHz. The horizontal axis is normalized by half of the TA mode wavelength at $f_\mathrm{FMR}$, $\tilde{\lambda}/2=\tilde{c}_t/2f_\mathrm{FMR}=375$ nm. When varying $d$ while keeping $L+d$ constant we observe phonon-pumping increased linewidths at the resonances indicated by vertical white lines [the resonance labeled (c2) in Fig. \ref{fig_MHzAnticrossing}].
On the other hand, for thicknesses $d=\tilde{\lambda},2\tilde{\lambda},\dots$, the magnetic precession is out of phase with the phonons and the coupling is suppressed.

When $L+d=0.5\ (\mathrm{mm})<\delta/2$, clear standing waves form by wave interference and the phonon spectrum is discrete [Fig. \ref{fig_MHzAnticrossing}(b)]. In addition to the resonance (c2), equidistant satellite peaks appear at the acoustic eigenfrequencies (c3) indicated by the white ladder on the ordinate. The effect of sound waves decreases for higher-order stress-matching conditions $m=2,3,\dots$ because of the interference with the Kittel mode, but may couple stronger to higher order perpendicular spin wave modes (R. Schlitz, private communication).
We observe clear avoided crossings of the FMR with the acoustic resonance frequencies when the three resonance conditions are simultaneously fulfilled. In Fig. \ref{fig_MHzAnticrossing}(b) we observed dips when the $f_\mathrm{FMR}=5.129$ GHz is tuned to the acoustic mode $n=1438$ and $m=1,2,3$. In Fig. \ref{fig_MHzAnticrossing}(d) we adopt the layer thicknesses of the sample used by An \textit{et al}. \cite{An2020} [red arrow in Fig. \ref{fig_MHzAnticrossing}(b)] and sweep the external field for the out-of-plane (left) and in-plane (right) magnetizations. The Kittel mode in the in-plane configuration is shifted to lower fields because of the thin-film shape anisotropy.
The anticrossing in the left panel indicates strong coupling between the magnetic and elastic excitations, as observed \cite{An2020}. Our estimates of the Kittel frequencies are slightly shifted from observed ones, which we tentatively attribute to residual anisotropies not included in our model. The sample thickness in the experiment deviates from the optimal stress-matching condition $m=1$, but the strong coupling is still achieved because of the broadness of the $m=1$ resonance, which is effective for a wider range of the YIG thickness $0.5\lesssim d/(0.5\tilde{\lambda})\lesssim1.5$. Note that the thin-film interference condition labeled (c4) [Eq. (\ref{ThinFilm})] favors the half-integers $d/(0.5\tilde{\lambda})=0.5, 1.5, \dots$.
On the right panel we again observe regularly spaced anticrossings, suggesting strong coupling between the Kittel and TA modes. The gap is, however, smaller compared to the normal configuration due to the absence of one of the transverse stresses, as shown in Fig. \ref{fig_AngleDependence} ($\sigma_\mathrm{me}^{yz}=0$ for $\theta_m=90^\circ$). Physically, the phonon pumping is less efficient because the emitted sound waves are now linearly polarized \cite{Streib2018}. The magnetoelastic stress does not rotate but oscillates and no net angular momentum is pumped into the NM.
Since Galfenol and GaAs have larger elastic damping, we choose for Fig. \ref{fig_MHzAnticrossing}(e) a thin NM film with $L=0.15$ mm ($<\delta/2 = 0.26$ mm). The spectra are broad due to the large Gilbert damping of Galfenol, yet exhibiting interaction with discrete TA modes for the $\theta_m=0^\circ$ configuration (left). When $\theta_m=90^\circ$ (right), they are not resolved because the large thin-film shape anisotropy significantly confines the precession within the film, suppressing $\sigma_\mathrm{me}^{xz}$ [see Eq. (\ref{sigma_me_1D_om}) and Fig. \ref{fig_AngleDependence}(c)].

The longitudinal waves interact with the dynamic magnetization when $0^\circ<\theta_m<90^\circ$. The maximum coupling is not universal but depends on the material parameters, found at $\theta_m=30^\circ$ for YIG$\vert$GGG and at $\theta_m=9^\circ$ for Galfenol$\vert$GaAs (Fig. \ref{fig_AngleDependence}). Fig. \ref{fig_LAAnticrossing}(a,c) shows the resonance conditions and (b,d) the FMR spectra at the magnetic film thickness indicated with the red arrows on the left.
Fig. \ref{fig_LAAnticrossing}(b) exhibits LA mode anticrossings. TA phonon features are suppressed because the thickness $d\sim513$ nm of the $m_l=1$ resonance lies in the middle between $m_t=1$ and 2 resonances, so the TA mode destructively interfere with the Kittel mode [see Fig. \ref{fig_MHzAnticrossing}(b)]. The selective coupling is possible in YIG$\vert$GGG bilayers because the transverse sound velocity $\tilde{c}_t$ in YIG is about a half of $\tilde{c}_\ell$. The anticrossing gap is smaller, however, since it is governed by the longitudinal MEC parameter $b_1\sim0.5b_2$.
Fig. \ref{fig_LAAnticrossing}(d) exhibits avoided crossings with not only TA but also LA phonons. At the simultaneous crossing with $n_t=320$ and $n_l=226$ modes the gap is large. We conclude that the Kittel mode can couple to the faster pressure waves for appropriate magnetization orientation and film thickness.

\section{\label{sec_disk} Applications: Magnetic grain}
Next we apply our formalism to a thin magnetic disk of radius $a$ embedded in a nonmagnetic film. The extension to, e.g., a spherical magnetic grain in a nonmagnetic matrix is straightforward. We focus on the regimes in which the film thickness is much smaller than the wavelengths of sound at frequencies up to several GHz, so the deformation is constant over the film thickness (in $z$-direction). We consider an infinitely extended medium, which means that emitted waves are not coming back. The MEC-BCs on the top and bottom surfaces of the film only rigidly shift the spectra and are therefore disregarded.

\begin{figure*}
    \centering
    \includegraphics[width=17.2cm]{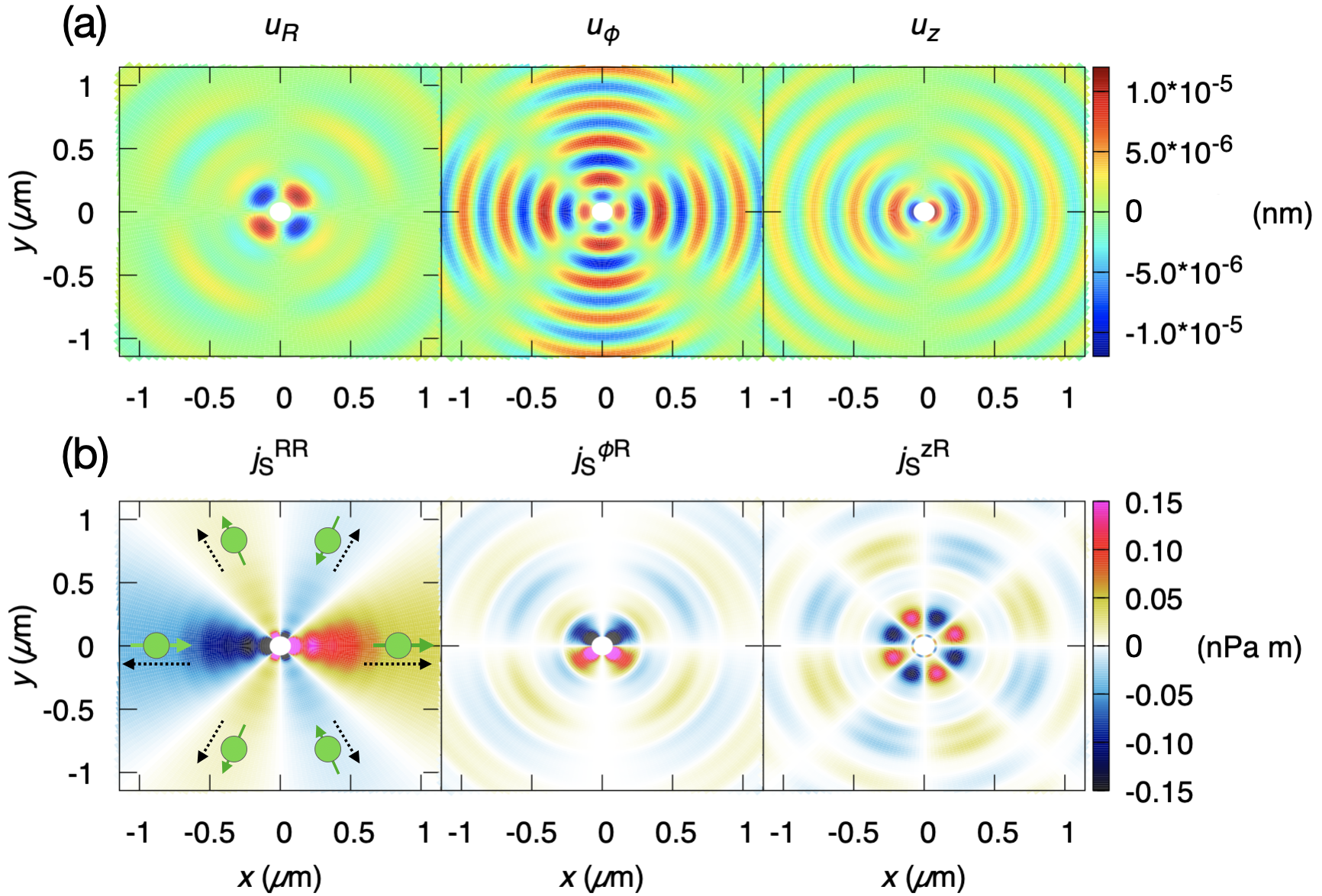}
    \caption{\label{fig_disk} Spatial profile of the (a) displacement components and (b) phonon spin currents at the FMR frequency 12.6 GHz emitted from a YIG disk of radius $a=76.2$ nm at the center (white). Magnetization precesses around the positive $x$-axis. Illustrations in (b), left, indicate the phonon spin orientation and propagation directions.}
\end{figure*}

The dipolar interaction of thin magnetic films favors in-plane magnetization. We define the $x$ axis along an in-plane external static magnetic field and introduce a cylindrical coordinate system: $x=R\cos\phi$ and $y=R\sin\phi$. Elastic waves are excited coherently by uniform magnetization precession
\begin{equation}\label{LLG_cylinder}
    \begin{pmatrix}
        m_y \\ m_z
    \end{pmatrix}
    =\mathsf{\chi}_\mathrm{FMR}(\omega,\pi/2)
    \begin{pmatrix}
        h_y \\ h_z
    \end{pmatrix},
\end{equation}
where the susceptibility tensor is defined in Eq. (\ref{pureFMR}) and $h_{y,z}$ are the magnetic fields of an applied microwave field at frequency $\omega$. The effect of strain in Eq. (\ref{LLG_1}) represents the back-action from the lattice and lead to the anticrossing spectra in Sec. \ref{sec_planar}. Here we focus on the propagation of pumped sound waves and disregard the higher-order self-consistent tickle fields in Eq. (\ref{LLG_cylinder}).
The displacement field in terms of scalar and vector displacement potentials read \cite{Graff}
\begin{equation}\label{Helmholtz_cylinder}
    \mathbf{u}(R,\phi,t)=\nabla\Phi(R,\phi,t)+\nabla\times\bm{\Xi}(R,\phi,t).
\end{equation}
Eq. (\ref{EoM_p}) separates into dilatation and shear motions with wave equations (in the absence of acoustic damping)
\begin{equation}\label{EOM_cylinder}
    \nabla^2\Phi=\ddot{\Phi}/c_\ell^2, \qquad
    \nabla^2\bm{\Xi}=\ddot{\bm{\Xi}}/c_t^2.
\end{equation}
The in- and out-of-plane motions decouple, which in the cylindrical coordinate system leads to independent pair of solutions $(\Phi,\Xi_z)$ and $(\Xi_R,\Xi_\phi)$, respectively, given by the Bessel functions multiplied by sinusoidal angular dependence (Appendix \ref{app_cylindrical}):
\begin{subequations}\label{Potentials_cylinder}
    \begin{align}
        \Phi&=
        \left\{
            \begin{aligned}
                \tilde{A}_\ell J_2(\tilde{k}_\ell R)\sin{2\phi} \\
                A_\ell H_2^{(1)}(k_\ell R)\sin{2\phi}
            \end{aligned}
        \right. \\
        \Xi_z&=
        \left\{
            \begin{aligned}
                \tilde{A}_tJ_2(\tilde{k}_tR)\cos{2\phi} \\
                A_tH_2^{(1)}(k_tR)\cos{2\phi}
            \end{aligned}
        \right. \\
        \Xi_R&=
        \left\{
            \begin{aligned}
                \tilde{C}J_0(\tilde{k}_tR)\sin{\phi} \\
                CH_0^{(1)}(k_tR)\sin{\phi}
            \end{aligned}
        \right. \\
        \Xi_\phi&=
        \left\{
            \begin{aligned}
                \tilde{C}J_0(\tilde{k}_tR)\cos{\phi} \\
                CH_0^{(1)}(k_tR)\cos{\phi}
            \end{aligned}
        \right.
    \end{align}
\end{subequations}
Here the upper rows hold for $0<R<a$ (in M) and the second for $a<R$ (in NM). $J_n$ and $H_n^{(1)}$ are the Bessel and Hankel functions of the first kind, respectively. For this model, the BCs (\ref{elBC_jp}) become
\begin{equation}\label{BCs_cylinder}
    \left\{
        \begin{aligned}
            u_\alpha(a-0,\phi)&=u_\alpha(a+0,\phi) \\
            \sigma_\mathrm{el}^{\alpha R}(a-0,\phi)+\sigma_\mathrm{me}^{\alpha R}(a-0,\phi)&=\sigma_\mathrm{el}^{\alpha R}(a+0,\phi)
        \end{aligned}
    \right.,
\end{equation}
where $\phi\in[0,2\pi]$, $\alpha=R,\phi,z$ and
\begin{equation}\label{sigma_me_cylinder}
    \begin{pmatrix}
        \sigma_\mathrm{me}^{RR} \\ \sigma_\mathrm{me}^{\phi R} \\ \sigma_\mathrm{me}^{zR}
    \end{pmatrix}
    =\begin{pmatrix}
        b_2m_y\sin{2\phi} \\ b_2m_y\cos{2\phi} \\ b_2m_z\cos\phi
    \end{pmatrix}.
\end{equation}
The angular dependence in Eq. (\ref{Potentials_cylinder}) follows from the constraint that Eq. (\ref{BCs_cylinder}) holds for arbitrary $\phi$. Eq. (\ref{BCs_cylinder}) then yields two decoupled matrix equations for the coefficients:
\begin{widetext}
    \begin{subequations}\label{BCs_cylinder_matrix}
        \begin{align}
            \begin{pmatrix}
                \tilde{A}_\ell  \\ A_\ell  \\ \tilde{A}_t \\ A_t
            \end{pmatrix}
            &=b_2a^2m_y
            \begin{pmatrix}
                \tilde{k}_\ell aJ_2^\prime(\tilde{k}_\ell a) & -k_\ell aH_2^{(1)\prime}(k_\ell a) & -2J_2(\tilde{k}_ta) & 2H_2^{(1)}(k_ta) \\
                2J_2(\tilde{k}_\ell a) & -2H_2^{(1)}(k_\ell a) & -\tilde{k}_taJ_2^\prime(\tilde{k}_ta) & k_taH_2^{(1)\prime}(k_ta) \\
                -\tilde{M}_\ell^R & M_\ell^R & \tilde{M}_t^R & -M_t^R \\
                -\tilde{M}_\ell^\phi & M_\ell^\phi & -\tilde{M}_t^\phi & M_t^\phi
            \end{pmatrix}^{-1}
            \begin{pmatrix}
                0 \\ 0 \\ 1 \\ 1
            \end{pmatrix}, \label{BCs_cylinder_matrix1}\\
            \begin{pmatrix}
                \tilde{C} \\ C
            \end{pmatrix}
            &=b_2a^2m_z
            \begin{pmatrix}
                \tilde{k}_taJ_0^\prime(\tilde{k}_ta) & -k_taH_0^{(1)\prime}(k_ta) \\
                -\tilde{\mu}(\tilde{k}_ta)^2J_0^{\prime\prime}(\tilde{k}_ta) & \mu(k_ta)^2H_0^{(1)\prime\prime}(k_ta)
            \end{pmatrix}^{-1}
            \begin{pmatrix}
                0 \\ 1
            \end{pmatrix}, \label{BCs_cylinder_matrix2}
        \end{align}
    \end{subequations}
    of in- and out-of-plane oscillations, respectively. The matrix elements
    \begin{equation}
        \begin{aligned}
            \tilde{M}_\ell^R&=2\tilde{\mu}(\tilde{k}_\ell a)^2J_2^{\prime\prime}(\tilde{k}_\ell a)-\tilde{\lambda}(\tilde{k}_\ell a)^2J_2(\tilde{k}_\ell a), &
            \tilde{M}_t^R&=4\tilde{\mu}\left[\tilde{k}_taJ_2^\prime(\tilde{k}_ta)-J_2(\tilde{k}_ta)\right], \\
            \tilde{M}_\ell^\phi&=4\tilde{\mu}\left[\tilde{k}_\ell aJ_2^\prime(\tilde{k}_\ell a)-J_2(\tilde{k}_\ell a)\right], &
            \tilde{M}_t^\phi&=\tilde{\mu}\left[2\tilde{k}_taJ_2^\prime(\tilde{k}_ta)+(\tilde{k}_ta)^2J_2(\tilde{k}_ta)\right],
        \end{aligned}
    \end{equation}
\end{widetext}
depend on frequency. The other components are given by removing tildes and replacing $J_2$ with $H_2^{(1)}$.

Eqs. (\ref{BCs_cylinder_matrix}) and (\ref{LLG_cylinder}) determine the phonon pumping. We consider a YIG nano-disk embedded in a GGG thin film. For the in-plane static field $\mu_0H_0=0.3609$ T and transverse microwave field $h_y=h_z=5$ A/m, the magnetization amplitudes are $|m_y(\omega)|\approx0.05$ and $|m_z(\omega)|\approx0.04$ at the FMR frequency $\sqrt{\omega_H(\omega_H+\omega_M-\omega_K)}/(2\pi)=12.6$ GHz. The disk radius $a=76.2$ nm satisfies the first stress-matching condition of the TA modes at this frequency.
The numerical solutions of Eqs. (\ref{BCs_cylinder_matrix}) for the displacement field at resonance are depicted in Fig. \ref{fig_disk}(a). The anisotropy of the magnetoelastic stress [Eq. (\ref{sigma_me_cylinder})] generates the observed angular patterns. The symmetry and location of the nodes do not depend on the material parameters. The magnetization precession (\ref{LLG_cylinder}) introduces a phase shift of $\pi/2$ between $u_\phi$ and $u_z$, i.e., excites rotational lattice motion in the vicinity of the $x$ axis.
In the absence of damping, the elastic energy of the wavefront $\propto\int_0^{2\pi}\mathbf{u}(R,\phi,\omega)^2d\phi$, decays as $1/R$ in the far field region by geometrical spreading, so $u_z$ decreases as $1/\sqrt{R}$. The amplitudes of $u_R$ and $u_\phi$ are coupled according to Eq. (\ref{BCs_cylinder_matrix1}), thereby oscillating as a function of $R$ by exchanging energy during propagation. The linear momentum current $j_p^{\alpha R}$ follows the angle dependence of $u_\alpha$ (not shown).

In Fig. \ref{fig_disk}(b) we plot the associated phonon spin current density in cylindrical coordinates (see Sec. \ref{sec_jS}),
\begin{equation}\label{jS_explicit}
    \left\{
        \begin{aligned}
            j_S^{RR}&=-u_zj_p^{\phi R}+u_\phi j_p^{zR} \\
            j_S^{\phi R}&=u_zj_p^{RR}-u_R j_p^{zR} \\
            j_S^{zR}&=-u_\phi j_p^{RR}+u_R j_p^{\phi R}
        \end{aligned}
    \right.,
\end{equation}
where the indices $\alpha$ and $\beta$ in $j_S^{\alpha\beta}$ refer to the phonon spin polarization and current direction, respectively. The plots in Fig. \ref{fig_disk}(b) represent the currents leaving the magnet in radial directions that vanish at the nodes of the respective displacement components, along which the sound waves are linearly polarized. Even though the displacement fields oscillate in time with the FMR frequency, $j_S^{RR}$ is a DC current, which is ensured by the $\pi/2$ phase shift and similar wavelengths between $u_\phi$ and $u_z$. The amplitude are extreme along the $x$-axis, transporting phonon spins $\parallel\hat{\mathbf{e}}_R$ in the forward angles $-\pi/4<\phi<\pi/4$ and spins $\parallel-\hat{\mathbf{e}}_R$ in the backward $3\pi/4<\phi<5\pi/4$ directions.
The other two components are AC currents carried dominantly by pressure waves $\Phi$.

\section{\label{sec_discussion} Discussion}
The stress tensors and the BCs derived in Sec. \ref{sec_formalism} are applicable to a wide range of materials of any crystal symmetry and MEC as well as geometries.
We focus in Secs. \ref{sec_planar} and \ref{sec_disk} on the Kittel mode excited by FMR in the presence of uniaxial crystalline and dipolar shape anisotropies. We can adopt MEC generated, e.g., by exchange \cite{Gurevich} or interfacial Dzyaloshinskii-Moriya interaction \cite{Xu2020,Kuess2020} to address phonon pumping by spin waves and magnetorotation coupling in basically any material combination with ferromagnets.
In principle, the analysis can also be extended to antiferromagnets \cite{Eremenko,Sasaki2019,Verba2019,Simensen2019,Li2020}, but the MEC energy and its parameters are less established. The magnon frequencies in antiferromagnets are typically higher than that of ultrasound \cite{Gurevich,Simensen2019} and magnon-phonon hybridization requires application of large magnetic fields \cite{Li2020}. Large MEC coefficients and large magnetization amplitudes are important for phonon pumping. Efficient phonon pumping does not necessarily require matching of wavenumbers as demonstrated in Secs. \ref{sec_planar} and \ref{sec_disk}, which unlocks the possibility of magnon-phonon strong coupling away from the intersections of their dispersion branches. More research is required to identify the best material combinations for an optimal sound generation by magnetization dynamics.

The BCs derived here allows deploying textbook knowledge of elastic waves \cite{Achenbach,Viktorov,Graff} to address various boundary shapes.
In Sec. \ref{sec_planar} we computed the phonon pumping in bilayers. M$\vert$NM$\vert$M phononic spin valves \cite{An2020} can be calculated by attaching another YIG layer to the free surface of GGG.
The transmission and reflection of sound waves in the opposite sandwich, i.e., a thin magnetic film inserted in an infinite nonmagnetic matrix \cite{Latcham2019}, is a simple extension of our model as a function of magnetization angle.
Our BCs can also address the energy partition between surface and bulk modes \cite{Lamb1904,Miller1954,Miller1955}. A magnetic stripline attached to a nonmagnetic substrate \cite{An2020,Zhang2020,Kei2020} excites not only the bulk phonons addressed here but also surface modes. Nonplanar structures such as acoustic whispering gallery modes around a sphere \cite{Rayleigh1910,Yamanaka2000,Sturman2015,Yamazaki2020}, or 3D ferromagnetic nanoparticles embedded in nonmagnetic media, as well as evanescent acoustic waves at interfaces with meta-materials \cite{Bliokh2019}, are within the scope of our formalism. Multiple magnets in a nonmagnetic matrix indirectly coupled via ultrasounds is a playground to study collective magnonic excitations, viz. a phononic extension of spin cavitronics \cite{CMHu2018,Nakamura2019}.

A large magnetostriction constant should lead to efficient phonon pumping, but may also induce a static deformation in the ground state that depends on the magnetization direction. For example, a magnetization $\mathbf{m}=(1,0,0)$ in the circular YIG disk discussed in Sec. \ref{sec_disk} generates static magnetoelastic stresses
\begin{equation}\label{sigma_st}
    \begin{pmatrix}
        \sigma_\mathrm{me}^{RR} \\ \sigma_\mathrm{me}^{\phi R} \\ \sigma_\mathrm{me}^{zR}
    \end{pmatrix}_\mathrm{static}
    =
    \begin{pmatrix}
        b_1\cos^2\phi \\
        -b_1\sin\phi\cos\phi \\
        0
    \end{pmatrix},
\end{equation}
that compresses the disk in the $x$ direction via the BC, $(\sigma_\mathrm{el}^{\alpha R}+\sigma_\mathrm{me}^{\alpha R})\vert_\mathrm{M}=\sigma_\mathrm{el}^{\alpha R}\vert_\mathrm{NM}$. This is consistent to the conventional static magnetostriction \cite{Kittel1949}. Other components of the magnetoelastic stress are finite as well, but the strains vanish by symmetry. The correction (\ref{sigma_st}) modify the phonon dispersion and the elastic constants \cite{Biot1940}. The correction $\sigma_\mathrm{me}\sim10^5$ Pa is, however, much smaller than the Lam\'{e} parameters $\sim10^{11}$ Pa in YIG, justifying that we disregard this effect in Secs. \ref{sec_planar} and \ref{sec_disk}.

In Sec. \ref{sec_disk} we discuss phonon pumping in elastically isotropic GGG, where the angular pattern solely originates from the anisotropy of $\sigma_\mathrm{me}$.
In other single crystals, the elastic anisotropy may further affect the angular dependence.

We focus here on microwave absorption experiments such as carried out by An \textit{et al}. \cite{An2020}. However, all experiments sensitive to the magnon polarons in YIG, such as pump and probe spectroscopy \cite{Ogawa2015,Hashimoto2018}, local and non-local spin Seebeck effect \cite{Kikkawa2016,Oyanagi2020}, and Brillouin light scattering \cite{Bozhko2017} are affected by the phonon pumping into GGG substrates and can in principle test our results.

\section{\label{sec_conclusion} Summary}
We present the BCs of magnet-nonmagnet composite systems for arbitrary interface geometries, magnetization orientation, and magnetoelastic interactions, with a focus on the consequences of MEC.
Our formalism is tuned to FMR conditions or small magnets, in which spin waves have little effect on the lattice and boundary dynamics becomes important. This natural extension of continuum mechanics allows transfer of knowledge from ultrasonics for a better understanding of the spintronics with phonons.

The phonon pumping scheme formulated here allows us to magnetically activate phonon modes in nonmagnets. Magnetic elements and fields may therefore be a tool to study quantum ground states of phonons and phononic computing \cite{Sklan2015}. The magnetization-angle- and geometry dependence of phonon pumping may be useful for engineering magnon-photon-phonon hybrids \cite{Zhang2016,Li2018,Nakamura2019}.
A formulation of the dynamics of magnetic core/shell type nanoparticles levitated in traps may require extension of our BCs to include effects of rigid body rotations \cite{Keshtgar2017,Andreas2020PRB} and oscillations \cite{Gonzalez2020}.

\acknowledgements
We thank Kei Yamamoto for fruitful discussions and sharing his insights.
The work was supported by JSPS KAKENHI Grants Nos. 20K14369 and 19H00645.
S. S. acknowledges financial support from the Knut and Alice Wallenberg Foundation through Grant No. 2018.0060.

\appendix
\section{\label{app_stress}Definition of stress tensors}
Our definition of the stress tensor (\ref{def_sigma_initial}) deviates from the conventional definition $\sigma_\mathrm{old}^{\alpha\beta}=\partial \mathcal{U}/\partial\varepsilon_{\alpha\beta}$, where $\varepsilon$ is the strain tensor. The latter only holds for infinite media with vanishing surface stresses \cite{Landauel}, i.e., disregarding surfaces and boundaries. The derivation of $\sigma_\mathrm{old}$ also postulates the symmetry of the stress tensor to ensure angular momentum conservation  \cite{Landauel,Graff}. In the presence of spin-lattice coupling, magnetization can be a source of angular momentum either in the bulk or at the boundaries, rendering the stress tensor asymmetric \cite{Tiersten1964,Tiersten1965,Garanin2015}, which is not reflected in $\sigma_\mathrm{old}$. The stress tensors defined here from the Lagrangian overcome these issues \cite{Tiersten1964,Tiersten1965,Akhiezer,Garanin2015}.

We can clarify the relation between $\sigma$ and $\sigma_\mathrm{old}$ by rewriting Eq. (\ref{def_sigma_initial}) in terms of strain and rotation tensors,
\begin{equation}\label{EpsilonOmega}
    \begin{aligned}
        \varepsilon_{\alpha\beta}(\mathbf{r},t)&=\frac{1}{2}(\partial_\beta u_\alpha+\partial_\alpha u_\beta), \\
        \omega_{\alpha\beta}(\mathbf{r},t)&=\frac{1}{2}(\partial_\beta u_\alpha-\partial_\alpha u_\beta),
    \end{aligned}
\end{equation}
i.e., by switching from the nine variables $\{\partial_\beta u_\alpha\}$ to the set of independent tensor elements, $\vec{\varepsilon}=(\varepsilon_{xx}, \varepsilon_{yy}, \varepsilon_{zz}, \varepsilon_{yz}, \varepsilon_{zx}, \varepsilon_{xy}, \omega_{yz}, \omega_{zx}, \omega_{xy})^T$. The first six components determines the elastic potential energy (\ref{def_U_el}) \cite{Landauel}, whereas the coupling (\ref{def_U_me}) in general depends on all nine elements. The chain rule leads to
\begin{widetext}
\begin{subequations}\label{def_sigma_matrix}
    \begin{alignat}{2}
        \sigma_\mathrm{el}
        &=
        \begin{pmatrix}
            \frac{\partial}{\partial\varepsilon_{xx}} & \frac{1}{2}\frac{\partial}{\partial\varepsilon_{xy}} & \frac{1}{2}\frac{\partial}{\partial\varepsilon_{zx}} \\
            \frac{1}{2}\frac{\partial}{\partial\varepsilon_{xy}} & \frac{\partial}{\partial\varepsilon_{yy}} & \frac{1}{2}\frac{\partial}{\partial\varepsilon_{yz}} \\
            \frac{1}{2}\frac{\partial}{\partial\varepsilon_{zx}} & \frac{1}{2}\frac{\partial}{\partial\varepsilon_{yz}} & \frac{\partial}{\partial\varepsilon_{zz}}
        \end{pmatrix}\mathcal{U}_\mathrm{el}, \label{def_sigmael}\\
        \sigma_\mathrm{me}
        &=
        \begin{pmatrix}
            \frac{\partial}{\partial\varepsilon_{xx}} & \frac{1}{2}\left(\frac{\partial}{\partial\varepsilon_{xy}}+\frac{\partial}{\partial\omega_{xy}}\right) & \frac{1}{2}\left(\frac{\partial}{\partial\varepsilon_{zx}}-\frac{\partial}{\partial\omega_{zx}}\right) \\
            \frac{1}{2}\left(\frac{\partial}{\partial\varepsilon_{xy}}-\frac{\partial}{\partial\omega_{xy}}\right) & \frac{\partial}{\partial\varepsilon_{yy}} & \frac{1}{2}\left(\frac{\partial}{\partial\varepsilon_{yz}}+\frac{\partial}{\partial\omega_{yz}}\right) \\
            \frac{1}{2}\left(\frac{\partial}{\partial\varepsilon_{zx}}+\frac{\partial}{\partial\omega_{zx}}\right) & \frac{1}{2}\left(\frac{\partial}{\partial\varepsilon_{yz}}-\frac{\partial}{\partial\omega_{yz}}\right) & \frac{\partial}{\partial\varepsilon_{zz}}
        \end{pmatrix}\mathcal{U}_\mathrm{me}. \label{def_sigmame}
    \end{alignat}
\end{subequations}
\end{widetext}
When calculating stress from $\sigma_\mathrm{old}$, differentiations with respect to off-diagonal strain components yield twice the correct value, as pointed out in Ref. \cite{Landauel}. This is because $\varepsilon_{\alpha\beta}\leftrightarrow\varepsilon_{\beta\alpha}$ and $\omega_{\alpha\beta}\leftrightarrow-\omega_{\beta\alpha}$ in the energy densities are not independent. Not taking care of the degrees of freedom of the strain tensor leads to a relation inconsistent with Eq. (\ref{def_sigmael}):
\begin{align}
    \sigma_\mathrm{el}^{\alpha\beta}&=\frac{\partial\mathcal{U}_\mathrm{el}}{\partial(\partial_\beta u_\alpha)}
    =\frac{\partial\varepsilon_{\nu\eta}}{\partial(\partial_\beta u_\alpha)}\frac{\partial\mathcal{U}_\mathrm{el}}{\partial\varepsilon_{\nu\eta}} \notag\\
    &=\frac{1}{2}(\delta_{\nu\beta}\delta_{\eta\alpha}+\delta_{\eta\beta}\delta_{\nu\alpha})\frac{\partial\mathcal{U}_\mathrm{el}}{\partial\varepsilon_{\nu\eta}} \notag\\
    &=\frac{\partial\mathcal{U}_\mathrm{el}}{\partial\varepsilon_{\alpha\beta}},
\end{align}
where the summation over $\nu,\eta$ doubly adds the off-diagonal elements. In our expression (\ref{def_sigma_matrix}), in contrast, the factor $1/2$ appropriately compensates the doubled values in the off-diagonal elements.
Eq. (\ref{def_sigmael}) implies that in the linear regime the elastic stress tensor is symmetric for any crystals whose elastic energy has the form (\ref{def_U_el}). Eq. (\ref{def_sigmame}) reveals that the symmetric part of the magnetoelastic stress arises from magnetostriction (coupling to strain), whereas its anti-symmetric part originates from the magnetorotation coupling. This implies that $\sigma_\mathrm{me}$ can differ from the conventional form $\partial\mathcal{U}_\mathrm{me}/(\partial\varepsilon_{\alpha\beta})$ when the magnetorotation coupling is relevant. In YIG \cite{Gurevich} and other ferromagnets such as Galfenol \cite{Clark2003,Parkes2013}, iron, and nickel \cite{Kittel1949} the effects of crystalline anisotropy are orders of magnitude smaller than that from the magnetostriction. However, the magnetorotation coupling is significant in CoFeB or Ni/Ag films that are thinner than acoustic wavelengths \cite{Xu2020,Puebla2020}.

\section{\label{app_1D} Average strain in 1D problem}
The linearized LLG reads [Eq. (\ref{LLG_1})]
\begin{equation}\label{LLG_1_app}
    \begin{pmatrix}
        m_\parallel \\ m_\perp
    \end{pmatrix}
    (\omega)
    =\mathsf{\chi}_\mathrm{FMR}(\omega,\theta_m)
    \left[
        \begin{pmatrix}
            h_\parallel \\ h_\perp
        \end{pmatrix}
        -\frac{1}{\gamma\mu_0}\begin{pmatrix}
            \Omega_\mathrm{me}^{\prime 13} \\ \Omega_\mathrm{me}^{\prime 23}
        \end{pmatrix}
    \right](\omega),
\end{equation}
where $\chi_\mathrm{FMR}$ reflects the purely magnetic response. We first compute the average strain in the magnetic film induced by MEC-BCs. For the one-dimensional problem, the components of the tensor $\mathsf{\Omega}_\mathrm{me}^\prime $ reads
\begin{equation}\label{Omega_prime}
    \begin{aligned}
        \Omega_\mathrm{me}^{\prime 13}&=\omega_c\partial_zu_x \cos{2\theta_m}-\omega_c^\ell \partial_zu_z\sin{2\theta_m}, \\
        \Omega_\mathrm{me}^{\prime 23}&=\omega_c\partial_zu_y \cos{\theta_m},
    \end{aligned}
\end{equation}
where $\omega_c=\omega_M/2+\gamma (b_2-K_1)/M_s$ and $\omega_c^\ell =\gamma b_1/M_s$ parameterize the magnetostriction and magnetorotation coupling. The BCs determine the relation between the complex acoustic wave amplitudes and magnetization. We then find the average strain in the magnet from Eq. (\ref{1D_Ansatz}) and write it in terms of magnetoelastic stress:
\begin{equation}\label{relative_displ_1D}
    \begin{aligned}
        \frac{u_\alpha(0,\omega)-u_\alpha(-d,\omega)}{d}
        &=-\frac{\sigma_\mathrm{me}^{\alpha z}(\theta_m)}{d\tilde{\rho}\tilde{c}_t}\frac{F(\omega)}{\omega+i\tilde{\eta}_\mathrm{el}/2},  \\
        \frac{u_z(0,\omega)-u_z(-d,\omega)}{d}
        &=-\frac{\sigma_\mathrm{me}^{zz}(\theta_m)}{d\tilde{\rho}\tilde{c}_\ell}\frac{F^\ell(\omega)}{\omega+i\tilde{\eta}_\mathrm{el}/2},
    \end{aligned}
\end{equation}
where $\alpha=x,y$ and
\begin{equation}
    F(\omega)=\frac{F_3+iF_4}{F_1+iF_2}, \qquad
    F^\ell(\omega)=\frac{F_3^\ell +iF_4^\ell }{F_1^\ell +iF_2^\ell }.
\end{equation}
The real-valued functions are defined as
\begin{widetext}
    \begin{equation}
        \begin{alignedat}{2}
            F_1&=&-\omega&(\beta_{11} \sin{\tilde{k}_td}\cos{k_tL}+\beta_{22} \cos{\tilde{k}_td}\sin{k_tL})
            +\frac{\tilde{\eta}_\mathrm{el}}{2}(\beta_{21}^\prime  \cos{\tilde{k}_td}\cos{k_tL}-\beta_{12}^\prime  \sin{\tilde{k}_td}\sin{k_tL}), \\
            F_2&=&\quad -\omega&(\beta_{21} \cos{\tilde{k}_td}\cos{k_tL}-\beta_{12} \sin{\tilde{k}_td}\sin{k_tL})
            -\frac{\tilde{\eta}_\mathrm{el}}{2}(\beta_{11}^\prime  \sin{\tilde{k}_td}\cos{k_tL}+\beta_{22}^\prime  \cos{\tilde{k}_td}\sin{k_tL}), \\
            F_3&=&\omega&[(\tilde{C}C+\beta_{11}) \cos{\tilde{k}_td}\cos{k_tL}-(\tilde{S}S+\beta_{22})\sin{\tilde{k}_td}\sin{k_tL}-2C\cos{k_tL}] \notag\\
            &\quad& +\frac{\tilde{\eta}_\mathrm{el}}{2}&[(\tilde{S}C+\beta_{21}^\prime )\sin{\tilde{k}_td}\cos{k_tL}+(\tilde{C}S+\beta_{12}^\prime )\cos{\tilde{k}_td}\sin{k_tL}-2S\sin{k_tL}], \\
            F_4&=&-\omega&[(\tilde{S}C+\beta_{21})\sin{\tilde{k}_td}\cos{k_tL}+(\tilde{C}S+\beta_{12})\cos{\tilde{k}_td}\sin{k_tL}-2S\sin{k_tL}] \notag\\
            &\quad& +\frac{\tilde{\eta}_\mathrm{el}}{2}&[(\tilde{C}C+\beta_{11}^\prime )\cos{\tilde{k}_td}\cos{k_tL}-(\tilde{S}S+\beta_{22}^\prime )\sin{\tilde{k}_td}\sin{k_tL}-2C\cos{k_tL}],
        \end{alignedat}
    \end{equation}
\end{widetext}
where
\begin{equation}
    \begin{aligned}
        \beta_{11}&=\tilde{C}C+\frac{\rho c_t}{\tilde{\rho}\tilde{c}_t}\tilde{S}S, & \beta_{12}&=\tilde{C}S+\frac{\rho c_t}{\tilde{\rho}\tilde{c}_t}\tilde{S}C, \\
        \beta_{21}&=\tilde{S}C+\frac{\rho c_t}{\tilde{\rho}\tilde{c}_t}\tilde{C}S, & \beta_{22}&=\tilde{S}S+\frac{\rho c_t}{\tilde{\rho}\tilde{c}_t}\tilde{C}C.
    \end{aligned}
\end{equation}
$C=\cosh{\kappa_t L}$ and $S=\sinh{\kappa_t L}$ represent the wave attenuation in NM, while $\tilde{C}=\cosh{\tilde{\kappa}_td}$ and $\tilde{S}=\sinh{\tilde{\kappa}_td}$ quantify the attenuation in M. $\beta_{ij}^\prime $ are given by multiplying $\eta_\mathrm{el}/\tilde{\eta}_\mathrm{el}$ to the impedance ratios.
$F^\ell(\omega)$ in Eq. (\ref{relative_displ_1D}) is defined with corresponding longitudinal parameters.

Replacing the strains in Eq. (\ref{Omega_prime}) with the average (\ref{relative_displ_1D}), and substituting the result into Eq. (\ref{LLG_1_app}) give Eq. (\ref{LLG_renorm}) in the main text.
In the limit of vanishing acoustic damping $\tilde{\eta}_\mathrm{el},\eta_\mathrm{el}\to0$, $\beta$ is diagonal, $F_2,F_4\to0$, and consequently the coupling strength [Eq. (\ref{def_g})] becomes real, i.e., magnetization damping is not enhanced by phonon pumping. The theory by Streib \textit{et al}. \cite{Streib2018} corresponds to the case $\tilde{\eta}_\mathrm{el}=0,\eta_\mathrm{el}\neq0$ and $L\to\infty$.

\section{\label{app_cylindrical} Elastic waves in a disk}
The Helmholtz relation between the displacement vector and potentials in cylindrical coordinates reads \cite{Graff}
\begin{equation}\label{u-PhiXi}
    \left\{
        \begin{aligned}
            u_R&=\frac{\partial \Phi}{\partial R}+\frac{1}{R}\frac{\partial \Xi_z}{\partial\phi}-\frac{\partial \Xi_\phi}{\partial z} \\
            u_\phi&=\frac{1}{R}\frac{\partial\Phi}{\partial\phi}-\frac{\partial\Xi_z}{\partial R}+\frac{\partial \Xi_R}{\partial z} \\
            u_z&=\frac{1}{R}\left[\frac{\partial}{\partial R}\left(R\Xi_\phi\right)-\frac{\partial \Xi_R}{\partial\phi}\right]+\frac{\partial\Phi}{\partial z}
        \end{aligned}
    \right.,
\end{equation}
where the $z$-derivatives vanish for thin films. The choice of the gauge $\mathrm{div}\bm{\Xi}=\psi(\mathbf{r},t)$, is a constraint on the four components $(\Phi,\Xi_\alpha)$ so that the both sides of Eq. (\ref{u-PhiXi}) have the same degrees of freedom \cite{Graff}. While $\psi(\mathbf{r},t)=0$ is suitable for planar configurations, we chose a different one for the present system as discussed below. The elastic strain tensor components read \cite{Graff}
\begin{subequations}
    \begin{align}
        \varepsilon_{RR}
        &=\frac{\partial u_R}{\partial R}, \\
        \varepsilon_{\phi\phi}
        &=\frac{1}{R}\frac{\partial u_\phi}{\partial\phi}+\frac{u_R}{R}, \\
        \varepsilon_{\phi z}
        &=\frac{1}{2R}\frac{\partial u_z}{\partial\phi}, \\
        \varepsilon_{zR}
        &=\frac{1}{2}\frac{\partial u_z}{\partial R}, \\
        \varepsilon_{R\phi}
        &=\frac{1}{2}\left(\frac{1}{R}\frac{\partial u_R}{\partial\phi}+\frac{\partial u_\phi}{\partial R}-\frac{u_\phi}{R}\right).
    \end{align}
\end{subequations}
The stress tensors transform as $\sigma=\mathcal{R}_z^{-1}\sigma_\mathrm{car} \mathcal{R}_z$, where $\sigma_\mathrm{car}$ is the tensor in Cartesian coordinates and the matrix $\mathcal{R}_z$ rotates the axes around the $z$ axis,
\begin{equation}
    \mathcal{R}_z(\phi)=\begin{pmatrix}
        \cos\phi & -\sin\phi & 0 \\
        \sin\phi & \cos\phi & 0 \\
        0 & 0 & 1
    \end{pmatrix}.
\end{equation}
The stress-strain relation takes the same form as in Cartesian coordinate systems since, using the property of the trace $\mathrm{tr}[\varepsilon_\mathrm{car}]=\mathrm{tr}[\mathcal{R}_z^{-1}\varepsilon_\mathrm{car}\mathcal{R}_z]$,
\begin{align}
    \sigma_\mathrm{el}^{\alpha\beta}
    &=(\mathcal{R}_z^{-1}\sigma_\mathrm{el,car}\mathcal{R}_z)_{\alpha\beta} \notag\\
    &=\lambda\delta_{\alpha\beta}\mathrm{tr}[\varepsilon_\mathrm{car}]+2\mu(\mathcal{R}_z^{-1}\varepsilon_{\mathrm{car}}\mathcal{R}_z)_{\alpha\beta} \notag\\
    &=\lambda\delta_{\alpha\beta}\mathrm{tr}[\varepsilon]+2\mu\varepsilon_{\alpha\beta},
\end{align}
for $\alpha,\beta\in\{R,\phi,z\}$. We then write the elastic stress within NM in terms of displacement potentials:
\begin{subequations}\label{sigma_el_cylinder}
    \begin{align}
        \sigma_\mathrm{el}^{RR}
        &=\left(2\mu\partial_R^2-\lambda k_\ell^2\right)\Phi+\frac{2\mu}{R}\partial_\phi\left(\partial_R-\frac{1}{R}\right)\Xi_z, \\
        \sigma_\mathrm{el}^{\phi R}
        &=\mu\left(\frac{2}{R}\partial_R+k_t^2\right)\Xi_z+\frac{2\mu}{R}\partial_\phi\left(\partial_R-\frac{1}{R}\right)\Phi, \\
        \sigma_\mathrm{el}^{zR}
        &=\mu\left[
            -\frac{\partial_\phi}{R}\left(\partial_R-\frac{1}{R}\right)\Xi_R
            \right. \notag\\
        &\qquad\left.
            +\left(\partial_R^2+\frac{1}{R}\partial_R-\frac{1}{R^2}\right)\Xi_\phi
        \right].
    \end{align}
\end{subequations}
The magnetoelastic stress tensor transforms analogously as $\sigma_\mathrm{me}=\mathcal{R}_z^{-1}\sigma_{\mathrm{me,car}}\mathcal{R}_z$. The components relevant for the BCs are, to linear order in transverse magnetization and for $\omega>0$,
\begin{equation}\label{app_sigma_me_cylinder}
    \begin{pmatrix}
        \sigma_\mathrm{me}^{RR} \\ \sigma_\mathrm{me}^{\phi R} \\ \sigma_\mathrm{me}^{zR}
    \end{pmatrix}
    =\begin{pmatrix}
        b_2m_y\sin{2\phi} \\ b_2m_y\cos{2\phi} \\ b_2m_z\cos\phi
    \end{pmatrix}.
\end{equation}
Magnetoelastic stress on a circular boundary does not depend on the longitudinal coupling $b_1$. In frequency domain the EOM for the deformation potentials (\ref{EOM_cylinder}) become
\begin{equation}\label{EoM_cylinder1}
    \left\{
        \begin{aligned}
            \left[\partial_R^2+\frac{1}{R}\partial_R+\frac{1}{R^2}\partial_\phi^2+k_\ell^2\right]&\Phi=0, \\
            \left[\partial_R^2+\frac{1}{R}\partial_R+\frac{1}{R^2}\partial_\phi^2+k_t^2\right]&\Xi_z=0, \\
            \left[\partial_R^2+\frac{1}{R}\partial_R+k_t^2+\frac{1}{R^2}\left(\partial_\phi^2-1\right)\right]&\Xi_R-\frac{2}{R^2}\frac{\partial\Xi_\phi}{\partial\phi}=0, \\
            \left[\partial_R^2+\frac{1}{R}\partial_R+k_t^2+\frac{1}{R^2}\left(\partial_\phi^2-1\right)\right]&\Xi_\phi+\frac{2}{R^2}\frac{\partial\Xi_R}{\partial\phi}=0,
        \end{aligned}
    \right.
\end{equation}
where $k_\ell =\omega/c_\ell$, $k_t=\omega/c_t$. Eqs. (\ref{sigma_el_cylinder}), (\ref{EoM_cylinder1}), and (\ref{u-PhiXi}) with $\partial_z=0$ confirm that the in-plane and out-of-plane dynamics decouple. By the separation of variables $\Phi(R,\phi)=f(R)X(\phi)$ and $\Xi_\alpha(R,\phi)=f_\alpha(R)X_\alpha(\phi)$, the first two equations in Eq. (\ref{EoM_cylinder1}) reduce to the Bessel differential equations for the radial component, whose solutions are given either by a linear superposition of the Bessel function of the first and second kind, $AJ_m(kR)+BY_m(kR)$, or by a combination of the Hankel function of the first and second kind, $AH_m^{(1)}(kR)+BH_m^{(2)}(kR)$. Since the amplitudes must be finite in M and backward waves represented by $H_m^{(2)}$ are not excited in the infinite NM,
\begin{align}
    f(R)&=
    \left\{
        \begin{aligned}
            &\tilde{A}_\ell J_m(\tilde{k}_\ell R) \\
            &A_\ell H_m^{(1)}(k_\ell R)
        \end{aligned}
    \right., \label{Ansatz_cylinder_f}\\
    f_z(R)&=
    \left\{
        \begin{aligned}
            &\tilde{A}_t J_n(\tilde{k}_tR) \\
            &A_t H_n^{(1)}(k_tR)
        \end{aligned}
    \right., \label{Ansatz_cylinder_fz}
\end{align}
where the first cases are for M $(0<R<a)$ and the second for NM $(a<R)$. $X(\phi)$ is then a superposition of $\cos{m\phi}$ and $\sin{m\phi}$ and $X_z(\phi)$ a superposition of $\cos{n\phi}$ and $\sin{n\phi}$, where $m,n$ are integers.
It follows from Eqs. (\ref{sigma_el_cylinder})(\ref{app_sigma_me_cylinder}) that the BCs (\ref{BCs_cylinder}) in the main text hold for all $\phi$ only if
\begin{align}
    X,\frac{dX_z}{d\phi}&\propto\sin{2\phi}, \\
    \frac{dX}{d\phi},X_z&\propto\cos{2\phi},
\end{align}
and thus we may assume $X(\phi)=\sin{2\phi}$ and $X_z=\cos{2\phi}$.

We next consider out-of-plane oscillations. If the BCs are to hold for arbitrary $\phi$, $X_\phi$ and $dX_R/d\phi$ must be proportional to $\cos\phi$, allowing us to assume $X_R=\sin\phi$ and $X_\phi=\cos\phi$. Out-of-plane dynamics in Eq. (\ref{EoM_cylinder1}) then becomes
\begin{equation}
    \begin{aligned}
        \left[\partial_R^2+\frac{1}{R}\partial_R+k_t^2-\frac{2}{R^2}\right]f_R+\frac{2}{R^2}f_\phi&=0, \\
        \left[\partial_R^2+\frac{1}{R}\partial_R+k_t^2-\frac{2}{R^2}\right]f_\phi+\frac{2}{R^2}f_R&=0.
    \end{aligned}
\end{equation}
Addition and subtraction of these equations give the Bessel differential equations
\begin{equation}
    \begin{aligned}
        \left[\partial_R^2+\frac{1}{R}\partial_R+k_t^2\right](f_R+f_\phi)&=0, \\
        \left[\partial_R^2+\frac{1}{R}\partial_R+k_t^2-\frac{4}{R^2}\right](f_R-f_\phi)&=0.
    \end{aligned}
\end{equation}
The general solutions are written as
\begin{align}
    f_R+f_\phi&=
    \left\{
        \begin{aligned}
            &2\tilde{C}J_0(\tilde{k}_tR) \\
            &2CH_0^{(1)}(k_tR)
        \end{aligned}
    \right., \label{Ansatz_cylinder_f+}\\
    f_R-f_\phi&=
    \left\{
        \begin{aligned}
            &2\tilde{C}^\prime J_2(\tilde{k}_tR) \\
            &2C^\prime H_2^{(1)}(k_tR)
        \end{aligned}
    \right., \label{Ansatz_cylinder_f-}
\end{align}
where we again discarded the second kind of the Bessel and Hankel functions. Eqs. (\ref{Ansatz_cylinder_f})(\ref{Ansatz_cylinder_fz})(\ref{Ansatz_cylinder_f+})(\ref{Ansatz_cylinder_f-}) contain 8 coefficients to be determined. Choosing an appropriate gauge $\psi(\mathbf{r},t)$, we may set $\tilde{C}^\prime=C^\prime=0$ without loss of generality \cite{Graff}, obtaining
\begin{equation}
    f_R(R)=f_\phi(R)=
    \left\{
        \begin{aligned}
            &\tilde{C}J_0(\tilde{k}_tR) \\
            &CH_0^{(1)}(k_tR)
        \end{aligned}
    \right..
\end{equation}
The remaining 6 coefficients are determined by the 6 BCs (\ref{BCs_cylinder}).

\bibliography{Refs}

\end{document}